\documentclass[11pt,a4paper,final]{amsart}

\usepackage{amsmath}
\usepackage{paralist}
\usepackage{graphics}
\usepackage{graphicx}
\usepackage{epstopdf}
\usepackage[colorlinks=true]{hyperref}
\usepackage{amssymb}
\usepackage{amsthm}
\usepackage{color}
\usepackage{multirow}
\usepackage{booktabs}
\usepackage{bm}
\usepackage{mathrsfs}

\usepackage{algorithm}
\usepackage{algorithmicx}
\usepackage{algpseudocode}
\usepackage{amsthm}
\usepackage{float}
\usepackage{comment}
\usepackage{xcolor}

\usepackage{tabularx}

\begin{document}

\title[Adaptive Time Stepping and Dynamic Penalty for American Put Options under the Bates Model]{Adaptive Time Stepping and Dynamic Penalty for American Put Options under the Bates Model}

\author{Neda Bagheri Renani${}^{1}$}
\author{Daniel \v{S}ev\v{c}ovi\v{c}${}^{1}$}

\address{${}^{1}$ Department of Applied Mathematics and Statistics, Faculty of Mathematics Physics and Informatics, Comenius University, Mlynsk\'a dolina, 842 48, Bratislava, Slovakia. Corresponding author: {\tt neda.bagheri@fmph.uniba.sk, sevcovic@fmph.uniba.sk} }

\begin{abstract}
We develop a coupled adaptive framework for pricing American
put options under the Bates stochastic-volatility jump-diffusion
model. The spatial grid is kept fixed, while the time-step size
and penalty parameter are adjusted dynamically using separate
temporal- and obstacle-error indicators. With a fixed penalty,
adaptive time stepping reduces the accepted steps from 200 to 32
with nearly unchanged reported errors. The fully coupled
strategy reduces the mean wall-clock time by approximately
39\% relative to the fixed-time dynamic-penalty method,
at the expense of increased numerical error, illustrating
the resulting cost--accuracy trade-off.
At three common error thresholds, FT+DP has lower mean
wall-clock time than AT+DP, indicating that the baseline
time reduction does not establish improved efficiency
at comparable accuracy.

\end{abstract}

\maketitle

\section{Introduction}
\label{sec:introduction}

Pricing American options under stochastic volatility and jumps is
numerically challenging because the multidimensional diffusion,
nonlocal jump term, and early-exercise constraint must be treated
simultaneously. Under the Bates model \cite{Bates1996}, the pricing
problem leads to an obstacle partial integro-differential equation.
Common numerical approaches include penalty methods
\cite{Zvan1998,DHalluin2004,ForsythVetzal2002}, ADI and
operator-splitting techniques \cite{IntHout2010,BagheriHaghighi2017},
and IMEX treatments of jump-diffusion terms \cite{Salmi2014}.

Temporal resolution is particularly important near maturity because of
the nonsmooth payoff and the rapid variation of the early-exercise
boundary. Adaptive time stepping can concentrate computational effort
where finer resolution is required
\cite{Soderlind2002,Persson2010,VonSydow2015}. At the same time,
the accuracy of obstacle enforcement depends on the penalty strength,
and non-constant penalty parameters have been considered in
\cite{Clevenhaus2020}. This motivates a joint treatment of temporal
adaptation and penalty control.

In this work, we develop a coupled adaptive framework for American put
option pricing under the Bates model while keeping the spatial
discretization fixed. 
The methodological contribution is the
feedback coupling of temporal step-size selection
and obstacle-based penalty adjustment, with
separate acceptance criteria for temporal and
obstacle errors. This coupling is assessed against
fixed-time and fixed-penalty alternatives, without
claiming improved efficiency at matched accuracy.
The numerical study evaluates separate and
coupled controllers through refinement tests,
boundary and parameter sensitivity, higher-jump
scenarios, and matched-accuracy cost comparisons.
The remainder of the paper is organized as follows.
Section~\ref{sec:Bates SVJ Model} presents the Bates model, while
Section~\ref{sec:framework} describes the numerical approximation.
Sections~\ref{sec:adaptive-time} and~\ref{sec:dynamic_penalty}
introduce the adaptive time-step and dynamic penalty controllers,
respectively. Section~\ref{sec:coupled_algorithm} combines these
components into the coupled adaptive algorithm.
Section~\ref{sec:numerical_experiments} presents the numerical
experiments and performance assessment, and
Section~\ref{sec:conclusion} concludes the paper.

\section{Model for American Options under the Bates SVJ Model}
\label{sec:Bates SVJ Model}
We consider an American-style option with payoff $g(S)$ under the Bates stochastic volatility jump-diffusion model that incorporates price jumps
\cite{Bates1996,Heston1993,Merton1976JFE,ContTankov2004Book}.
The early-exercise feature of American options leads to a partial-integro differential inequality:
\begin{equation}\label{eq:american_obstacle_V}
\min\Bigl\{
-V_t(S,v,t)-\mathcal{A}[V](S,v,t)-\mathcal{I}[V](S,v,t),\;
V(S,v,t)-g(S)
\Bigr\}=0,
\end{equation}
for $(S,v,t)\in(0,\infty)\times(0,\infty)\times[0,T)$, with terminal condition $V(S,v,T)=g(S)$. Here, $V(S,v,t)$ denotes the value of an American put option with maturity $T$ and payoff $g(S)=(K-S)^+$, $K>0$ is the strike price. The local differential operator $\mathcal{A}$  and the nonlocal jump operator $\mathcal{I}$ are defined as:
\begin{equation*}
\begin{aligned}
\mathcal{A}[V](S,v,t)
&:= \frac{1}{2}S^2vV_{SS}
+\rho\sigma vSV_{Sv}
+\frac{1}{2}\sigma^2vV_{vv}
+rSV_S
+\kappa(\theta-v)V_v
-rV,
\\
\mathcal{I}[V](S,v,t)
&:=
\lambda \int_{\mathbb{R}}
\Bigl(
V(Se^z,v,t)-V(S,v,t)-(e^z-1)SV_S(S,v,t)
\Bigr)f(z)\,dz.
\end{aligned}
\label{eq:local_A} 
\end{equation*}
The compensator term ensures the integrability of the jump integral and preserves the martingale property of the discounted asset-price process \cite{Merton1976JFE,ContTankov2004Book}.
Setting $\lambda=0$ recovers the Heston model,
whereas constant variance reduces the above system to the Merton
jump-diffusion model
\cite{Bates1996,Heston1993,Merton1976JFE} with normal distribution of jumps. 

Recall that the variational inequality \eqref{eq:american_obstacle_V} with operators
$\mathcal{A}[V]$ and $\mathcal{I}[V]$ is derived according to standard principles, using the It\=o lemma and assuming that the stochastic processes $\{S_t,\; t\geq 0\}$ and $\{v_t,\; t\geq 0\}$ are stochastic processes driving the underlying asset price and instantaneous variance follow the system of stochastic differential equations involving jumps:
\begin{equation*}
\begin{aligned}
dS_t &= \mu S_t\,dt + \sqrt{v_t}\,S_t\,dW_t^S + S_t\,dJ_t,\\
dv_t &= \kappa(\theta-v_t)\,dt + \sigma\sqrt{v_t}\,dW_t^v,
\qquad dW_t^S\,dW_t^v=\rho\,dt.
\end{aligned}
\end{equation*}
where $\kappa$, $\theta$, $\sigma$, and $\rho$ are the mean-reversion speed,
long-term variance, variance volatility, and instantaneous correlation,
respectively \cite{Bates1996,Heston1993}. The jump process $J(t)$ is a compound Poisson process with intensity $\lambda>0$ and multiplicative jumps $S \mapsto S e^{z},$
where the logarithmic jump size $z$ has density $f(z)$ and is independent of the increments of the Wiener processes $\left(dW_t^{S},dW_t^{v}\right)$ that have correlation $\rho\in[-1,1]$ (see \cite{Merton1976JFE,ContTankov2004Book,Andersen2000}).
In the numerical experiments, the logarithmic jump size is assumed to satisfy
$z\sim\mathcal{N}(m_J,\delta_J^2)$.
The drift $\mu$ is chosen so that the discounted asset price is a martingale under the risk-neutral measure. 

Using standard transformations: $x=\ln(S/K), \tau=T-t, u(x,v,\tau)=V(S,v,t)/ K$ and 
$\widetilde{g}(x)=(1-e^x)^+$. In these variables, the problem is written as the obstacle PIDE:
\begin{equation*}
\min\left\{
u-\widetilde g,\;
u_\tau-\mathcal{L}u-\mathcal{J}u
\right\}=0,
\qquad
u(x,v,0)=\widetilde g(x).
\end{equation*}
where the  differential operator $\mathcal{L}$  and the nonlocal jump operator $\mathcal{J}$ are defined as:
\begin{equation*}
\begin{aligned}
\mathcal{L}u
&=
\frac{v}{2}\left(u_{xx}-u_x\right)
+\rho\sigma v\,u_{xv}
+\frac{\sigma^2v}{2}\,u_{vv}
+\kappa(\theta-v)u_v
+r u_x-r u\,,
\\
\mathcal{J}u
&=
\lambda
\int_{\mathbb{R}}
\left[
u(x+z,v,\tau)
-u(x,v,\tau)
-\left(e^z-1\right)u_x(x,v,\tau)
\right]f(z)\,dz\ .
\end{aligned}
\label{eq:bates-sde-transformed}
\end{equation*}

\section{Numerical approximation scheme}
\label{sec:framework}
For the variable temporal discretization we set 
\begin{equation*}
0=\tau_0<\tau_1<\cdots<\tau_N=T,
\qquad
h_n=\tau_{n+1}-\tau_n,
\qquad
U^n\approx u(\cdot,\cdot,\tau_n),
\end{equation*}
where the proposed steps satisfy $h_{\min}\leq h_n\leq h_{\max}$,
except for a possibly shorter final step imposed by \eqref{eq:final-step}.
 The spatial mesh and the spatial discretization are kept unchanged during the adaptive simulation. The local multidimensional part is addressed by the fixed spatial/ADI solver, whereas the nonlocal jump component is handled explicitly in the IMEX framework. Numerical approaches for the Bates SVJ model can be found in \cite{BagheriRenaniSevcovic2026}, and directional splitting techniques for option pricing are treated in \cite{IntHout2010,BagheriHaghighi2017}. IMEX schemes for jump-diffusion operators are studied in \cite{Salmi2014}.
For the sake of brevity, one numerical update over a proposed temporal
interval is represented by
\[
U^{n+1}
=
\mathcal{S}_{\mathrm{sp}}
\left(
U^n;h_n,\varepsilon_n
\right),
\]
where $\varepsilon_n>0$ is the penalty parameter. The notation $S_{\mathrm{sp}}$ suppresses the
dependence on the jump-history values and the preceding time-step size;
these quantities are explicitly specified in Section~\ref{sec:adaptive-time}.
The American option pricing obstacle problem is enforced through the penalty function
\[
\mathcal{P}_{\varepsilon_n}(U)
=
\frac{1}{\varepsilon_n}
(\widetilde g-U)^+.
\]
Smaller values of $\varepsilon_n$ correspond to stronger enforcement of the
obstacle.

In our paper, the penalty parameter is allowed to vary during
the calculation; its coupling to the temporal step size and its subsequent
correction according to the obstacle residual are developed in Section~\ref{sec:dynamic_penalty}.
The penalized problem is written in the form
\[
u_\tau-\mathcal{L}u-\mathcal{J}u
=
P_{\varepsilon_n}(u)
=
\frac{1}{\varepsilon_n}(\tilde g-u)^+.
\]
For a fixed $\varepsilon_n$ within a temporal trial, the penalized
problem is solved by a relaxed Picard iteration. In the Picard iteration $k$, let $U_{\mathrm{cand}}^{(k+1)}$ denote the
solution obtained with the penalty term frozen at the current iteration
$U^{(k)}$. The relaxed update is
\[
U^{(k+1)}
=
U^{(k)}
+
\omega
\left(
U_{\mathrm{cand}}^{(k+1)}-U^{(k)}
\right),
\]
where $\omega=0.75$. The iteration is terminated when
\[
\Vert
U_{\mathrm{cand}}^{(k+1)}-U^{(k)}
\Vert_{\infty}
\le 10^{-9},
\]
A maximum of 100 Picard iterations is allowed for each nonlinear problem solution.

The adaptive procedure uses two indicators: $\eta_n$ controls temporal acceptance, while $\zeta_n$ measures the normalized violation of the American obstacle. Their definitions and update rules are introduced  in Sections~\ref{sec:adaptive-time} and~\ref{sec:dynamic_penalty}, respectively.

The unbounded spatial domain is truncated to $\Omega=[x_{\min},x_{\max}]\times[0,v_{\max}]$. Since $1-e^{x_{\min}}\to 1$  as $x_{\min}\to-\infty$ and $u(+\infty,v,\tau)=0$ we impose the boundary conditions
\begin{equation*}
u(x_{\min},v,\tau)=1,
\qquad
u(x_{\max},v,\tau)=0.
\end{equation*}
At the finite boundary
$x_{\min}=-5$, the asymptotic value
$u=1$ differs from the exercise payoff
$1-e^{-5}$ by $e^{-5}\approx
6.738\times10^{-3}$. The left-exterior
jump extension instead uses
$1-e^{x+z}$ when $x+z<x_{\min}$.
The sensitivity to this mismatch is
examined in
Section~\ref{sec:variance_boundary_sensitivity}.

In the $v$ direction we assume Neumann type boundary conditions
\begin{equation*}
u_v(x,0,\tau)=0,
\qquad
u_v(x,v_{\max},\tau)=0,
\end{equation*}
with one-sided finite-difference closures at the boundary nodes.

At $v=0$, the coefficients of the
variance-diffusion and mixed-derivative terms
vanish, whereas the drift term
$\kappa\theta u_v$ remains nonzero in general.
Therefore, the homogeneous Neumann condition
at this degenerate boundary is a numerical
closure rather than a condition derived from
the limiting Bates equation.
For the baseline parameters,
$2\kappa\theta=0.16>\sigma^2=0.09$,
so the Feller condition is satisfied.
The influence of this numerical closure on the
computed option prices is examined in
Section~\ref{sec:variance_boundary_sensitivity}.

For evaluation of the jump integral, shifted points outside the
log-price interval are extended according to
\[
u^{\mathrm{ext}}(x+z,v,\tau)
=
\begin{cases}
1-e^{x+z}, & x+z<x_{\min},\\
0, & x+z>x_{\max}.
\end{cases}
\]

Within each solver run, the spatial grid, boundary treatment, and jump quadrature remain fixed. Thus, temporal adaptation
and penalty updates do not alter the
spatial discretization.

\subsection{Discrete spatial operators and Douglas ADI stages}
\label{subsec:discrete-douglas}

Let $x_i=x_{\min}+i\Delta x$ and $v_j=j\Delta v$, and store the
values $U_{j,i}$ with the variance index first. For either coordinate
$\xi\in\{x,v\}$ with mesh width $\Delta\xi$, the interior first- and
second-derivative approximations $d_i^{(1)}$ and $d_i^{(2)}$ satisfy the
fourth-order compact relations
\begin{equation*}
\begin{aligned}
\tfrac14d_{i-1}^{(1)}+d_i^{(1)}+\tfrac14d_{i+1}^{(1)}
 &=\frac{3(q_{i+1}-q_{i-1})}{4\Delta\xi},\\
\tfrac1{12}d_{i-1}^{(2)}+\tfrac{10}{12}d_i^{(2)}
 +\tfrac1{12}d_{i+1}^{(2)}
 &=\frac{q_{i-1}-2q_i+q_{i+1}}{(\Delta\xi)^2}.
\end{aligned}
\end{equation*}
At the lower endpoint, the derivative rows use the one-sided formulas
\begin{equation*}
\begin{aligned}
d_0^{(1)}&=\frac{-25q_0+48q_1-36q_2+16q_3-3q_4}
 {12\Delta\xi},\\
d_0^{(2)}&=\frac{45q_0-154q_1+214q_2-156q_3+61q_4-10q_5}
 {12(\Delta\xi)^2}.
\end{aligned}
\end{equation*}
The upper-endpoint rows use the reflected coefficients, with the sign
reversed for the first derivative. Solving the corresponding compact
linear systems defines $D_x,D_{xx},D_v,D_{vv}$.

The local discrete Bates operator is split as
$\mathcal L_h=\mathcal L_{x,h}+\mathcal L_{v,h}+\mathcal L_{0,h}$,
where, with $I_x$ and $I_v$ denoting the coordinate identity matrices,
\begin{equation*}
\begin{aligned}
L_x^{(j)}&=\tfrac12v_jD_{xx}
 +(r-\tfrac12v_j)D_x-\tfrac12rI_x,\\
L_v&=\operatorname{diag}(\tfrac12\sigma^2v_j)D_{vv}
 +\operatorname{diag}(\kappa(\theta-v_j))D_v-\tfrac12rI_v,\\
\mathcal L_{0,h}U&=\rho\sigma\operatorname{diag}(v_j)
 \bigl(D_vUD_x^{\mathsf T}\bigr).
\end{aligned}
\end{equation*}
Here $\mathcal L_{x,h}$ applies $L_x^{(j)}$ to each variance row, and
$\mathcal L_{v,h}U=L_vU$. Thus the reaction term $-ru$ is divided
equally between the two directional operators. The mixed derivative
$\mathcal L_{0,h}$ is evaluated explicitly. 

In the implementation,
$\mathcal{L}_{x,h}U$ is zeroed at the
$x$-boundary columns,
$\mathcal{L}_{v,h}U$ at both the
$v$-boundary rows and $x$-boundary
columns, and $\mathcal{L}_{0,h}U$
at all boundary nodes.

The outer $x$ rows of the directional systems replace the local PDE
with the prescribed Dirichlet equations; the outer $v$ rows of the
variance-direction system impose the one-sided homogeneous Neumann
relations from the one-sided relations above. The resulting
variance-boundary values are also reconstructed after each completed
Douglas step.

The Gaussian jump integral is truncated symmetrically at the grid-aligned
nodes $z_m=m\Delta x$, $-M\leq m\leq M$, where
$M=\max\{2,\lceil(|m_J|+8\delta_J)/\Delta x\rceil\}$.
With composite Simpson weights
$w_m=(\Delta x/3)c_mf(z_m)$, where $c_m=1$ at the endpoints,
$c_m=4$ when $m+M$ is odd and $c_m=2$ at the remaining interior nodes, the
implemented jump operator is
\begin{equation*}
(\mathcal J_hU)_{j,i}=\lambda\sum_{m=-M}^{M}w_m
\bigl[U^{\mathrm{ext}}_{j,i+m}-U_{j,i}
 -(e^{z_m}-1)(D_xU)_{j,i}\bigr].
\end{equation*}
Because every shift $z_m$ is a multiple of $\Delta x$, interior shifted
values coincide with grid nodes and require no off-grid interpolation.
Outside the interval, $U^{\mathrm{ext}}$ uses the extension prescribed
above; the jump contribution is zeroed at the two Dirichlet columns.
The reported prices, in contrast, are postprocessed by bilinear
interpolation in $(S/K,v)$.

For a trial step of length $h$ and a fixed penalty parameter
$\varepsilon_n$, let
$F^{n,(k)}=J_{\mathrm{AB2}}^{n+1}
 +\mathcal P_{\varepsilon_n}(U^{(k)})$,
with the penalty forcing set to zero at the Dirichlet columns.
The actual Douglas update, with $\vartheta=\tfrac12$, consists of
\begin{equation*}
\begin{aligned}
Y_0={}&U^n+h\bigl[\mathcal L_{x,h}U^n
 +\mathcal L_{v,h}U^n+\mathcal L_{0,h}U^n
 +F^{n,(k)}\bigr],\\
(I-\vartheta h\mathcal L_{x,h})Y_1
 ={}&Y_0-\vartheta h\mathcal L_{x,h}U^n,\\
(I-\vartheta h\mathcal L_{v,h})Y_2
 ={}&Y_1-\vartheta h\mathcal L_{v,h}U^n.
\end{aligned}
\end{equation*}
The first directional system is solved independently for each $v_j$
with Dirichlet rows at $x_{\min},x_{\max}$. The second is solved
independently for each interior $x_i$ with homogeneous Neumann rows at
$v=0,v_{\max}$. After the boundary values are enforced, the resulting
state is the Picard candidate $U_{\mathrm{cand}}^{(k+1)}$ used in the
relaxed iteration described earlier in this section. The jump forcing
uses the accepted-step history in \eqref{eq:variable-ab2}; the full-step
and half-step histories are specified in
Section~\ref{subsec:step-doubling}. The explicit mixed-derivative term
is not corrected by an additional Craig-Sneyd stage.

\section{Adaptive Time-Step Selection}
\label{sec:adaptive-time}
The temporal behavior of an American put can vary substantially over the
pricing interval. In particular, the nonsmooth payoff and the rapid movement
of the early-exercise boundary near expiration may require finer resolution close
to $\tau=0$. Numerical and analytical studies of this near-expiration behavior
are given in \cite{Stamicar1999,LaukoSevcovic2010}.

Adaptive time stepping provides a natural mechanism for concentrating
computational effort in regions where the solution changes rapidly, while
allowing larger steps once the solution becomes smoother. The general principles
of automatic step-size control are discussed in \cite{Soderlind2002}.
Adaptive finite difference methods for the pricing of American options are considered
in \cite{Persson2010}, while adaptive finite differences combined with the integration of IMEX
 for the Bates SVJ model are studied in \cite{VonSydow2015}.
The spatial grid remains fixed,
and each temporal error comparison uses
a fixed penalty; Section~\ref{sec:dynamic_penalty}
describes its dynamic update.
\subsection{Variable-Step IMEX Update}
\label{subsec:variable-imex}

For $n\geq1$, define the step-size ratio $\chi_n=h_n/h_{n-1}$. Using the discrete jump values introduced in Section~\ref{sec:framework}, the variable-step AB2 extrapolation is as follows.
\begin{equation}
J_{\mathrm{AB2}}^{n+1}
=
\left(1+\frac{\chi_n}{2}\right)J^n
-\frac{\chi_n}{2}J^{n-1}.
\label{eq:variable-ab2}
\end{equation}
For $\chi_n=1$, this reduces to the standard constant-step AB2 formula.
The dependence on the recent step-size ratio is consistent with
variable-step linear multistep formulations \cite{Arevalo2021}.

To examine the explicit jump contribution,
consider the scalar test equation $y'=\mu y$,
with $\mu=-a<0$. For a fixed step-size ratio
$\chi>0$, the variable-step AB2 recurrence
has both characteristic roots strictly
inside the unit disk when
$0<ah<2/(1+\chi)$.
For $\chi=1$, this reduces to the standard
AB2 restriction $ah<1$.
This frozen-ratio scalar condition does not
establish stability for arbitrary variable-step
sequences or for the coupled IMEX--ADI--penalty
scheme. Its implications are examined in
Section~\ref{subsec:jump-stability-diagnostic}.

One trial update applies the explicit variable-step AB2 jump
forcing \eqref{eq:variable-ab2} together with the Douglas predictor and
two implicit directional solves described in
Section~\ref{subsec:discrete-douglas}, following the Douglas ADI
(alternating direction implicit) framework \cite{IntHout2010,BagheriHaghighi2017}. The nonlocal jump term is explicit, as in IMEX treatments
of jump-diffusion problems \cite{Salmi2014}. The mixed derivative is
also evaluated explicitly, and the penalty term is frozen within each Douglas solve
and updated by the relaxed Picard iteration of
Section~\ref{sec:framework}. Consequently, the implemented split update
is not algebraically identical to an unsplit Crank--Nicolson equation;
in particular, choosing $\vartheta=1/2$ does not establish
second-order temporal convergence in the presence of an explicitly
treated mixed derivative. The variable-step jump forcing is interpreted
as an approximation to the jump contribution over the trial interval,
not as a computed value at the future time level.

Because the AB2 extrapolation requires two previous jump evaluations, a
start-up step is performed before adaptive control begins. At
$\tau_0=0$, only
\[
J^0=\mathcal{I}_{J}[U^0]
\]
is available, so the jump contribution is treated by an explicit-Euler
start-up using $J^0$ over the prescribed step $h_0$. After computing $U^1$, we set
\[
J^1=\mathcal{I}_{J}[U^1].
\]
The pair $(J^0,J^1)$ then provides the history required by
\eqref{eq:variable-ab2}. No temporal error indicator is assigned to the
start-up step.

\subsection{Step-Doubling Error Control and Step-Size Update}
\label{subsec:step-doubling}

Temporal adaptation compares one full step with two consecutive half
steps over the same interval:
\begin{equation}
U_F^{n+1}
=
\mathcal{S}_{\mathrm{sp}}
\left(
U^n;h_n,\varepsilon_n
\right),\ \ 
U_H^{n+\frac12} = \mathcal{S}_{\mathrm{sp}}
\left(
U^n;\frac{h_n}{2},\varepsilon_n
\right),\ \ 
U_H^{n+1} = \mathcal{S}_{\mathrm{sp}}
\left(
U_H^{n+\frac12};\frac{h_n}{2},\varepsilon_n
\right).
\label{eq:full-step}
\end{equation}
The same penalty value $\varepsilon_n$ is used in all three trial calculations. The corresponding AB2 ratios are
\[
\chi_F=\frac{h_n}{h_{n-1}},
\qquad
\chi_{H,1}=\frac{h_n/2}{h_{n-1}}.
\]
Before the second half step, the jump term is reevaluated as
\[
J_H^{n+\frac12}
=
\mathcal{I}_{J}
\left[
U_H^{n+\frac12}
\right],
\]
and the equal consecutive half steps give $\chi_{H,2}=1$.
The step-doubling discrepancy is
\begin{equation}
E_n
=
\frac{1}{3}
\left(
U_H^{n+1}-U_F^{n+1}
\right).
\label{eq:step-doubling-error}
\end{equation}

The factor $1/3$ in
\eqref{eq:step-doubling-error} follows the
classical step-doubling scaling for a
second-order one-step method. However,
this scaling is not theoretically established
for the present variable-step AB2--Douglas
ADI--penalty scheme. In particular, the
full-step and half-step computations use
different jump-history ratios. Therefore,
$E_n$ is treated as a practical temporal
indicator rather than a guaranteed local-error
estimate. Its numerical behavior is examined
in Section~\ref{subsec:experimental-setup}.

Next, the normalized temporal indicator is
\begin{equation}
\eta_n
=
\max_{i,j}
\frac{
|E_{n,i,j}|
}{
\mathrm{atol}
+
\mathrm{rtol}
\cdot
\max
\left(
|U_{H,i,j}^{n+1}|,
|U_{F,i,j}^{n+1}|
\right)
}.
\label{eq:eta}
\end{equation}
Mixed absolute-relative scaling follows standard adaptive
time-integration practice \cite{Soderlind2002}. A trial satisfies the temporal criterion when
$\eta_n\leq1$. 
In that case, $U_H^{n+1}$ is retained as the trial state and passed to
the obstacle controller of Section~\ref{sec:dynamic_penalty}; otherwise,
the trial is repeated with a smaller time step.

Let us define the cut-off function $\Pi_{[a,b]}(z)=\min\{b,\max\{a,z\}\}$.  With safety factors $0<s<1$ and $\eta_{\mathrm{floor}}>0$, the step-size
factor is
\begin{equation}
q_n
=
\Pi_{[q_{\min},q_{\max}]}
\left[
s
\left(
\frac{1}{
\max(\eta_n,\eta_{\mathrm{floor}})
}
\right)^{1/3}
\right],
\label{eq:q-controller}
\end{equation}
and the proposed step size is
\begin{equation}
h_{\mathrm{new}}
=
\Pi_{[h_{\min},h_{\max}]}
\left(
q_n h_n
\right).
\label{eq:h-new}
\end{equation}

The exponent $1/3$ in
\eqref{eq:q-controller} is retained as
an empirical step-size control parameter,
rather than a consequence of proven
second-order temporal accuracy.
Its use does not imply a rigorous
local-error bound for the complete scheme.

For an accepted trial, $h_{\mathrm{new}}$ provides the next proposed
step; for a rejected trial, it replaces the current step before the
calculation is repeated. Before each trial,
\begin{equation}
h_n
\leftarrow
\min\{h_n,T-\tau_n\},
\label{eq:final-step}
\end{equation}
so that the computation ends exactly at $\tau=T$. If the remaining
interval is smaller than $h_{\min}$, that remaining interval is used as
the final step.
If a trial has $\eta_n>1$ when
$h_n=h_{\min}$, further temporal refinement
is impossible under the prescribed step-size
bounds. The trial is rejected and the
computation terminates with a diagnostic
error rather than accepting a step that
violates the temporal tolerance. The same
safeguard applies when the final remaining
interval is shorter than $h_{\min}$ and
cannot be reduced further.

\section{Dynamic Penalty Parameter Selection}
\label{sec:dynamic_penalty}
The penalty operator has been introduced in Section~\ref{sec:framework}.
In this section we specify how its parameter $\varepsilon_n$ is
selected and updated during adaptive computation. The lower values of
$\varepsilon_n$ correspond to stronger enforcement of the American
obstacle.

Penalty methods for American option valuation and their numerical properties have been studied in several settings \cite{Zvan1998,DHalluin2004,ForsythVetzal2002,Wang2006,Zhang2008}, while non-constant penalty parameters are considered in \cite{Clevenhaus2020}. The penalty parameter is determined through two complementary mechanisms: a time-step-based candidate and a correction based on the
measured obstacle violation.

\subsection{Time-Step-Based Penalty and Obstacle Indicator}
\label{subsec:penalty-indicator}

For a proposed temporal step $h_n$, define the baseline penalty value by
\begin{equation}
\varepsilon_n^{\mathrm{time}}
=
\Pi_{[\varepsilon_{\min},\varepsilon_{\max}]}
\left(
c_\varepsilon h_n^\alpha
\right),
\qquad
c_\varepsilon>0,
\quad
\alpha>0 .
\label{eq:penalty-time}
\end{equation}
The coefficient $c_\varepsilon$ determines the overall penalty scale, while $\alpha$ controls its dependence on the current temporal resolution. In particular, $\alpha=1$ gives a linear dependence on $h_n$.
A simple motivation for the coupling between the temporal step size and
the penalty parameter can be obtained from an error-balancing argument.
Suppose, heuristically, that the temporal discretization error behaves as
$C_t h^p$, while the error associated with the penalization behaves as
$C_\varepsilon \varepsilon^q$ over the parameter range of interest.
Balancing these two contributions gives
\[
C_t h^p \approx C_\varepsilon \varepsilon^q,
\]
and hence suggests a scaling of the form
\[
\varepsilon \propto h^{p/q}.
\]
This motivates the general relation
$\varepsilon_n^{\mathrm{time}}=c_\varepsilon h_n^\alpha$ used here,
with $\alpha$ acting as a tunable coupling exponent. We do not interpret
this relation as an asymptotically exact error law for the full Bates
obstacle problem; rather, it provides a practical mechanism by which the
penalty scale is adjusted consistently with the current temporal
resolution. The sensitivity experiments in Section~\ref{subsec:penalty-sensitivity} examine the
effect of varying $c_\varepsilon$ and $\alpha$ around their baseline
values.

At the initial level, no previous obstacle information or previously
accepted penalty value is available, so the parameter is initialized by
$\varepsilon_0=\varepsilon_0^{\mathrm{time}}$.

After a temporal trial satisfies the acceptance criterion
$\eta_n\leq1$, the retained trial state is checked against the American
obstacle. Define
\begin{equation}
R_n^{\mathrm{obs}}
=
\Vert
(\widetilde g-U_{\mathrm{trial}}^{n+1})^+
\Vert_\infty,
\qquad
\zeta_n
=
\frac{R_n^{\mathrm{obs}}}
{\mathrm{tol}_{\mathrm{pen}}},
\label{eq:obstacle-indicator}
\end{equation}
where $\mathrm{tol}_{\mathrm{pen}}>0$ is the prescribed obstacle tolerance. The obstacle requirement is satisfied when
\begin{equation}
\zeta_n\leq1.
\label{eq:obstacle-criterion}
\end{equation}
Thus, $\eta_n$ controls temporal acceptance, whereas $\zeta_n$ measures obstacle enforcement. The two indicators are evaluated separately because satisfying the temporal criterion does not by itself guarantee that the American obstacle is enforced to the prescribed tolerance.
Since $R_{\mathrm{obs}}$ measures only
obstacle feasibility, we also examine the
PIDE supersolution inequality and
complementarity in
Section~\ref{subsec:complementarity-diagnostics}.
These additional diagnostics do not alter
the controller acceptance rules.

Since the option value is normalized by
the strike, the baseline obstacle tolerance
$\mathrm{tol}_{\mathrm{pen}}=0.005$ corresponds to
a maximum permitted gridwise shortfall of $0.5$ price units from the exercise payoff for $K=100$. This tolerance does not guarantee absolute pricing accuracy.

\subsection{Within-Step Penalty Correction}
\label{subsec:within-step-penalty}

Suppose that a trial satisfies the temporal criterion but violates the
obstacle condition \eqref{eq:obstacle-criterion}. The temporal step is
initially retained, while the penalty is strengthened according to
\begin{equation*}
\varepsilon_n
\leftarrow
\max
\left\{
\varepsilon_{\min},
\gamma_{\mathrm{dec}}\varepsilon_n
\right\},
\qquad
0<\gamma_{\mathrm{dec}}<1.
\label{eq:penalty-strengthening}
\end{equation*}

Because changing $\varepsilon_n$ changes the penalized solution, the
full-step and two-half-step calculations are repeated using the updated
penalty parameter. The same updated value is used in the full-step and
both half-step calculations so that the temporal comparison remains
internally consistent. The temporal indicator $\eta_n$ and obstacle
indicator $\zeta_n$ are then reevaluated.

If $\varepsilon_n$ has already reached $\varepsilon_{\min}$ and the
obstacle requirement remains unsatisfied, temporal refinement is used as
a safeguard:
\begin{equation}
h_n
\leftarrow
\max
\left\{
h_{\min},
\gamma_\tau h_n
\right\},
\qquad
0<\gamma_\tau<1.
\label{eq:penalty-time-safeguard}
\end{equation}

Whenever the current trial step $h_n$ is reduced before acceptance,
whether because of temporal rejection, the obstacle safeguard above, or
failure of the nonlinear Picard solve, the baseline penalty is
recomputed from \eqref{eq:penalty-time}. For adaptive steps after
start-up, the recomputed value is additionally restricted relative to
the penalty value from the previously accepted step by the factors
$q_{\varepsilon,\min}$ and $q_{\varepsilon,\max}$. This prevents an
abrupt change in penalty strength when a trial is repeated with a new
time step.

If a nonlinear Picard solver fails and $h_n>h_{\min}$, the time step is
reduced according to \eqref{eq:penalty-time-safeguard}, the corresponding
penalty value is recomputed, and the trial is restarted. If nonlinear
convergence cannot be obtained at $h_{\min}$, the computation is
terminated.

Likewise, if both lower bounds
\[
h_n=h_{\min},
\qquad
\varepsilon_n=\varepsilon_{\min}
\]
have been reached while
$\zeta_n>1$, the prescribed obstacle tolerance cannot be achieved under
the selected parameter bounds, and the computation is terminated rather
than continuing an indefinite correction loop.

\subsection{Penalty Update After an Accepted Step}
\label{subsec:penalty-update}
Once a trial satisfies both $\eta_n\leq1$ and $\zeta_n\leq1$, 
the state is accepted and the temporal controller proposes the next step
size $h_{n+1}$. The corresponding time-based penalty value is obtained
from \eqref{eq:penalty-time} with $h_n$ replaced by $h_{n+1}$.

To avoid unnecessarily strong penalization when the obstacle residual is
well below its prescribed tolerance, introduce
\[
0<\theta_{\mathrm{pen}}<1
\]
and define the provisional penalty value
\begin{equation}
\widehat{\varepsilon}_{n+1}
=
\begin{cases}
\gamma_{\mathrm{inc}}\,
\varepsilon_{n+1}^{\mathrm{time}},
&
\zeta_n<\theta_{\mathrm{pen}},
\\[1mm]
\varepsilon_{n+1}^{\mathrm{time}},
&
\theta_{\mathrm{pen}}\leq\zeta_n\leq1,
\end{cases}
\qquad
\gamma_{\mathrm{inc}}>1.
\label{eq:penalty-relaxation}
\end{equation}

Hence, the penalty may be moderately relaxed when the obstacle residual is substantially smaller than the requested tolerance. No relaxation is applied when the residual is closer to the acceptance threshold.
Finally, let us choose
\[
0<q_{\varepsilon,\min}\leq1,
\qquad
q_{\varepsilon,\max}\geq1.
\]
The proposed penalty value for the next step is restricted by both
absolute and relative bounds:
\begin{equation}
\varepsilon_{n+1}
=
\Pi_{
\left[
\max\left(
\varepsilon_{\min},
q_{\varepsilon,\min}\varepsilon_n
\right),
\,
\min\left(
\varepsilon_{\max},
q_{\varepsilon,\max}\varepsilon_n
\right)
\right]
}
\left(
\widehat{\varepsilon}_{n+1}
\right).
\label{eq:penalty-next}
\end{equation}

Therefore, the update combines the temporal information contained in $h_{n+1}$ with the obstacle information contained in $\zeta_n$, while respecting both the absolute and relative bounds of the penalty parameter.
The complete interaction between the temporal and penalty controllers is assembled into a single algorithm in Section~\ref{sec:coupled_algorithm}.
\section{Coupled Adaptive Time-Stepping and Dynamic Penalty Algorithm}
\label{sec:coupled_algorithm}
The temporal controller developed in Section~\ref{sec:adaptive-time}
and the penalty controller of Section~\ref{sec:dynamic_penalty}
are now combined into a single adaptive procedure.
The spatial discretization, boundary treatment, and jump quadrature
introduced in Section~\ref{sec:framework} remain fixed throughout the
computation, so the coupled procedure modifies only the temporal step size
and the penalty parameter.

The temporal component follows standard adaptive step-size principles
\cite{Soderlind2002,Arevalo2021}, whereas the penalty mechanism is related
to the use of non-constant penalty parameters for American option valuation
\cite{Clevenhaus2020}. The particular coupling used here is determined by
the update rules developed in Sections~\ref{sec:adaptive-time}
and~\ref{sec:dynamic_penalty}.
The start-up procedure provides
$U^1$ and the AB2 jump history without
a temporal indicator, while retaining
obstacle and Picard safeguards.
Thereafter, each trial checks nonlinear
convergence, $\eta_n\leq1$, and
$\zeta_n\leq1$ sequentially. Failed
criteria invoke the applicable
step-size or penalty safeguards;
accepted steps update both controls.
Algorithm~\ref{alg:coupled-adaptive} summarizes these operations
using the equations of Sections~\ref{sec:adaptive-time} and~\ref{sec:dynamic_penalty}.

\begin{algorithm}[ht]
\caption{Coupled adaptive time-step and penalty procedure}
\label{alg:coupled-adaptive}

\footnotesize
\begin{algorithmic}[1]

\Require
$U^0$, $T$, $h_0$, $\mathrm{atol}$, $\mathrm{rtol}$,
$\mathrm{tol}_{\mathrm{pen}}$, and controller parameters

\State Set $\tau_0\gets0$
\State Initialize $\varepsilon_0$ from \eqref{eq:penalty-time}
\State Compute $J^0\gets\mathcal{I}_{J}[U^0]$

\State Compute and accept the start-up state $U^1$ using the
procedure of Section~\ref{subsec:variable-imex} and the safeguards of
Section~\ref{subsec:within-step-penalty}

\State Set $\tau_1\gets h_0$ and compute
$J^1\gets\mathcal{I}_{J}[U^1]$
\State Set $n\gets1$

\While{$\tau_n<T$}

    \State Restrict the proposed step using \eqref{eq:final-step}
    \State $\mathrm{accepted}\gets\mathrm{false}$

    \While{$\mathrm{accepted}=\mathrm{false}$}

        \State Compute the full-step and two-half-step solutions using
        \eqref{eq:full-step}

        \If{a nonlinear solver fails}
            \State Apply the safeguard of
            Section~\ref{subsec:within-step-penalty} and repeat the trial
        \Else

            \State Evaluate $E_n$ and $\eta_n$ using
            \eqref{eq:step-doubling-error} and \eqref{eq:eta}

        \If{$\eta_n>1$}
        \If{$h_n\le h_{\min}$}
           \State \textbf{Terminate} with a temporal-tolerance diagnostic
    \Else
          \State Update $h_n$ using
        \eqref{eq:q-controller} and \eqref{eq:h-new}
           \State Update the corresponding penalty according to
        Section ~\ref{subsec:within-step-penalty}
    \EndIf
\Else

                \State Set
                $U_{\mathrm{trial}}^{n+1}\gets U_H^{n+1}$
                \State Evaluate $R_n^{\mathrm{obs}}$ and $\zeta_n$
                using \eqref{eq:obstacle-indicator}

                \If{$\zeta_n>1$}
                    \State Apply the penalty correction or temporal
                    safeguard of Section~\ref{subsec:within-step-penalty}
                \Else
                    \State $\mathrm{accepted}\gets\mathrm{true}$
                \EndIf

            \EndIf

        \EndIf

    \EndWhile

    \State Accept
    $U^{n+1}\gets U_{\mathrm{trial}}^{n+1}$
    and set $\tau_{n+1}\gets\tau_n+h_n$

    \State Compute
    $J^{n+1}\gets\mathcal{I}_{J}[U^{n+1}]$

   \State Determine $h_{\mathrm{new}}$ using
   \eqref{eq:q-controller} and \eqref{eq:h-new}, and set
    $h_{n+1}\leftarrow h_{\mathrm{new}}$

    \State Update $\varepsilon_{n+1}$ using
    \eqref{eq:penalty-time},
    \eqref{eq:penalty-relaxation}, and \eqref{eq:penalty-next}

    \State $n\gets n+1$

\EndWhile

\Ensure Numerical approximation at $\tau=T$

\end{algorithmic}
\end{algorithm}
\section{Numerical Experiments and Performance Assessment}
\label{sec:numerical_experiments}
This section evaluates the numerical behavior of the temporal controller,
the dynamic penalty mechanism, and their coupled implementation developed
in Sections~\ref{sec:adaptive-time}--\ref{sec:coupled_algorithm}.

\subsection{Experimental Setup and Working Reference}
\label{subsec:experimental-setup}

We consider an American put option on the truncated computational domain
$\Omega=[-5,5]\times[0,1]$ using the fixed spatial grid
$(N_x,N_v)=(100,50)$. Unless stated otherwise, the model parameters and
the numerical control parameters are set to the baseline values listed
in Table~1. Parameters that are varied in individual experiments are
explicitly specified in the corresponding subsections.

All computations were performed on an ASUS ExpertBook laptop equipped
with an Intel Core i5 processor. Whenever wall-clock time is reported in
the computational-cost experiments, each configuration was executed
10 times in the same software and hardware environment. The reported
wall-clock time is the mean over the 10 runs, and the associated
variability is measured by the sample standard deviation. Therefore,
these timing results are used only for relative comparisons among the
numerical configurations tested.

Before carrying out the adaptive experiments, the implementation was
verified in fixed-step mode against the underlying fixed-step solver.
For
\[
(N_x,N_v,N_t)=(50,25,50)
\quad\text{and}\quad
(N_x,N_v,N_t)=(100,50,200),
\]
the two implementations agreed to machine precision.

Since no closed-form solution is available for the American option
pricing problem under the Bates model, a refined numerical solution is
used as a working reference. To avoid conflating temporal refinement
with penalty refinement, the two effects are examined separately on the
fixed spatial grid $(N_x,N_v)=(100,50)$.

First, the penalty parameter is fixed at
\[
\varepsilon=1.25\times10^{-3},
\]
while the temporal grid is successively refined according to
\[
N_t\in\{800,1600,3200\},
\qquad
\Delta\tau=\frac{T}{N_t}.
\]
For two successive temporal refinement levels, we define
\[
D_\infty^{(t)}(N_t,2N_t)
=
\left\|
U_{2N_t,\varepsilon}
-
U_{N_t,\varepsilon}
\right\|_\infty .
\]

Second, the temporal grid is fixed at
\[
N_t=3200,
\qquad
\Delta\tau=3.125\times10^{-4},
\]
while the penalty parameter is successively refined as
\[
\varepsilon\in
\left\{
1.25\times10^{-3},
6.25\times10^{-4},
3.125\times10^{-4}
\right\}.
\]
For two successive penalty levels, we define
\[
D_\infty^{(\varepsilon)}(\varepsilon,\varepsilon/2)
=
\left\|
U_{N_t,\varepsilon/2}
-
U_{N_t,\varepsilon}
\right\|_\infty .
\]

The resulting refinement values are reported in
Table~\ref{tab:reference-refinement}. The option values
$V(80)$, $V(100)$, and $V(120)$ are evaluated at $v=0.04$.
In the temporal-refinement experiment, the successive maximum-norm
differences decrease as $N_t$ is increased while the final obstacle
residual remains essentially unchanged. In the penalty-refinement
experiment, the successive maximum-norm differences decrease as
$\varepsilon$ is reduced, while the final obstacle residual decreases
simultaneously. These two experiments therefore provide separate
checks of stabilization with respect to the temporal discretization and
the penalty approximation.

The computation corresponding to
\[
N_t=3200,
\qquad
\varepsilon=3.125\times10^{-4},
\]
is used as the working numerical reference in the subsequent
experiments. It is not regarded as an exact solution. Moreover, because
the spatial grid remains fixed throughout the refinement procedure, the
reported errors do not represent complete space--time discretization
errors.

For a computed solution $U$, we measure the error in the maximum norm
$E_\infty$ and in the discrete $L^2$ norm evaluated at $\tau=T$:
\[
E_\infty=\|U-U_{\mathrm{ref}}\|_\infty,
\qquad
E_2=
\left[
h_xh_v
\sum_{i,j}
\left(
U_{i,j}-U_{\mathrm{ref},i,j}
\right)^2
\right]^{1/2}.
\]
We also report the maximum error in the selected option prices at
$v=0.04$:
\[
E_{\mathrm{price}}
=
\max_{S\in\{80,100,120\}}
\left|
V(S,0.04)-V_{\mathrm{ref}}(S,0.04)
\right|.
\]

Throughout the remainder of this section, $N_{\mathrm{acc}}$ and
$N_{\mathrm{rej}}$ denote the numbers of accepted time steps and
rejected temporal trials, respectively. ``Picard'' denotes the
cumulative number of Picard iterations performed during the
corresponding computation.

\begin{table}[ht]
\centering
\caption{\small Baseline model and numerical parameters.}
\label{tab:baseline_parameters}

\footnotesize
\small

\setlength{\tabcolsep}{3pt}
\renewcommand{\arraystretch}{1.12}

\begin{tabular}{cc|cc|cc}
\hline
Parameter & Value &
Parameter & Value &
Parameter & Value \\
\hline

$K$ & $100$
& $T$ & $1$
& $r$ & $0.05$ \\

$\kappa$ & $2.0$
& $\theta$ & $0.04$
& $\sigma$ & $0.30$ \\

$\rho$ & $-0.50$
& $\lambda$ & $0.05$
& $(m_J,\delta_J)$ & $(-0.05,0.10)$ \\

ADI scheme & Douglas
& $\theta_{\mathrm{ADI}}$ & $0.5$
& Jump term & AB2 \\

$h_0$ & $5\times10^{-3}$
& $[h_{\min},h_{\max}]$
& $[10^{-4},5\times10^{-2}]$
& $\mathrm{atol}$ & $10^{-5}$ \\

$\mathrm{rtol}$ & $10^{-3}$
& $s$ & $0.90$
& $\eta_{\mathrm{floor}}$ & $10^{-12}$ \\

$(q_{\min},q_{\max})$
& $(0.5,2)$
& $(c_\varepsilon,\alpha)$
& $(1,1)$
& $[\varepsilon_{\min},\varepsilon_{\max}]$
& $[10^{-8},10^{-1}]$ \\

$\theta_{\mathrm{pen}}$
& $0.1$
& $(\gamma_{\mathrm{dec}},\gamma_{\mathrm{inc}})$
& $(0.5,2)$
& $\gamma_\tau$
& $0.5$ \\

$(q_{\varepsilon,\min},q_{\varepsilon,\max})$
& $(0.25,4)$
& $\mathrm{tol}_{\mathrm{pen}}$
& $5\times10^{-3}$
& Picard tol.
& $10^{-9}$ \\

Picard relaxation
& $0.75$
& & &
& \\
\hline
\end{tabular}%
\end{table}
The baseline upper bound
$h_{\max}=0.05$ is an empirical controller
safeguard, not a proven stability limit
for the coupled variable-step IMEX--ADI
scheme. In particular, the explicit AB2
jump term has no separately enforced
spectral stability bound.
A scalar AB2 restriction and numerical
tests varying the jump intensity,
step-size cap, and step-size ratios are
reported in
Section~\ref{subsec:jump-stability-diagnostic}.

\begin{table}[H]
\centering
\caption{Separate temporal and penalty refinement tests on the fixed spatial grid $(N_x,N_v)=(100,50)$. In panel (a), the penalty parameter is fixed at
$\varepsilon=1.25\times10^{-3}$ while the temporal grid is refined.
In panel (b), the temporal grid is fixed at $N_t=3200$ while the penalty
parameter is refined. The quantity $D_\infty$ denotes the maximum-norm
difference between two successive refinement levels.}
\label{tab:reference-refinement}

\footnotesize
\setlength{\tabcolsep}{3.5pt}

\begin{tabular}{ccccccc}
\toprule
\multicolumn{7}{c}{\textbf{(a) Temporal refinement with fixed
$\varepsilon=1.25\times10^{-3}$}}\\
\midrule
$N_t$ & $\Delta\tau$ & $V(80)$ & $V(100)$ & $V(120)$ &
$R_{\mathrm{obs}}^{\mathrm{final}}$ & $D_\infty$\\
\midrule
800
& $1.2500\times10^{-3}$
& 19.995097
& 6.060612
& 1.894645
& $1.1128\times10^{-4}$
& $1.30\times10^{-6}$\\

1600
& $6.2500\times10^{-4}$
& 19.995099
& 6.060690
& 1.894701
& $1.1118\times10^{-4}$
& $6.13\times10^{-7}$\\

3200
& $3.1250\times10^{-4}$
& 19.995100
& 6.060723
& 1.894724
& $1.1113\times10^{-4}$
& --\\
\bottomrule
\end{tabular}

\vspace{2mm}

\begin{tabular}{ccccccc}
\toprule
\multicolumn{7}{c}{\textbf{(b) Penalty refinement with fixed $N_t=3200$}}\\
\midrule
$\varepsilon$ & $\Delta\tau$ & $V(80)$ & $V(100)$ & $V(120)$ &
$R_{\mathrm{obs}}^{\mathrm{final}}$ & $D_\infty$\\
\midrule
$1.2500\times10^{-3}$
& $3.1250\times10^{-4}$
& 19.995100
& 6.060723
& 1.894724
& $1.1113\times10^{-4}$
& $4.31\times10^{-5}$\\

$6.2500\times10^{-4}$
& $3.1250\times10^{-4}$
& 19.997554
& 6.061269
& 1.894935
& $6.8052\times10^{-5}$
& $2.17\times10^{-5}$\\

$3.1250\times10^{-4}$
& $3.1250\times10^{-4}$
& 19.998778
& 6.061538
& 1.895037
& $4.6394\times10^{-5}$
& --\\
\bottomrule
\end{tabular}
\end{table}

The temporal refinement results
in Table~\ref{tab:reference-refinement}(a)
give a ratio of approximately $2.119$
between successive maximum-norm differences,
corresponding to an observed order of
$1.083$. Thus, second-order temporal
convergence is not observed in this regime.
Additional step-doubling tests on a fixed
$(N_x,N_v)=(48,24)$ spatial grid, with
$\varepsilon=0.04$, showed that the
indicator \eqref{eq:step-doubling-error}
underestimated the half-step error relative
to a refined numerical reference by factors
of approximately $3.05$--$3.26$.
Tests along selected accepted adaptive steps
gave corresponding factors of
$3.28$--$3.76$.
These findings support the interpretation
of $E_n$ as an empirical temporal indicator,
without establishing a rigorous local-error
estimate or second-order accuracy.

We also examined the explicit-Euler jump
start-up of Section~\ref{subsec:variable-imex}
using a diagnostic predictor--corrector applied
only to the first step. On the fixed
$(N_x,N_v)=(48,24)$ grid with $\varepsilon=0.04$,
both start-up choices gave observed temporal
orders close to one. For the baseline adaptive
run with $h_0=0.005$, the change in $V(100,0.04)$
was $4.14\times10^{-7}$, with 35 accepted steps
for either choice. At $\lambda=0.20$, the
corresponding difference was
$1.63\times10^{-6}$. Thus, the explicit-Euler
jump start-up does not account for the observed
near-first-order temporal behavior in the
tested regimes.

\subsection{Adaptive Time-Step Behavior}
\label{subsec:adaptive-time-results}

We first examine the temporal controller independently by fixing $\varepsilon=0.04$. For this experiment, errors are measured against a matched fixed-penalty
reference computed on the same spatial grid with $N_t=800$ and
$\varepsilon=0.04$. This comparison therefore isolates temporal adaptation from changes in the penalty approximation.

We use $ \mathrm{atol}=10^{-5},\ \ 10^{-4}\leq h_n\leq5\times10^{-2}$. We consider
\[
\mathrm{rtol}
\in
\left\{
10^{-2},
10^{-3},
5\times10^{-4},
10^{-4}
\right\}.
\]
Table~\ref{tab:adaptive_time_results} summarizes the resulting adaptive
step sequences.
\begin{table}[ht]
\centering
\footnotesize

\setlength{\tabcolsep}{3pt}
\renewcommand{\arraystretch}{0.95}
\caption{Adaptive time-step results with fixed $\varepsilon=0.04$.}
\label{tab:adaptive_time_results}

\begin{tabular}{cccccccc}
\hline
$\mathrm{rtol}$ &
$N_{\mathrm{acc}}$ &
$N_{\mathrm{rej}}$ &
$h_{\min}^{\mathrm{used}}$ &
$\overline h$ &
$\eta_{\max}$ &
$E_\infty$ &
$E_{\mathrm{price}}$ \\
\hline

$10^{-2}$ &
29 & 1 &
$4.159\times10^{-3}$ &
$3.448\times10^{-2}$ &
0.9005 &
$5.156\times10^{-5}$ &
$3.148\times10^{-3}$ \\

$10^{-3}$ &
32 & 2 &
$3.168\times10^{-3}$ &
$3.125\times10^{-2}$ &
0.8032 &
$4.887\times10^{-5}$ &
$2.729\times10^{-3}$ \\

$5\times10^{-4}$ &
33 & 2 &
$2.983\times10^{-3}$ &
$3.030\times10^{-2}$ &
0.7987 &
$4.898\times10^{-5}$ &
$2.641\times10^{-3}$ \\

$10^{-4}$ &
36 & 2 &
$2.476\times10^{-3}$ &
$2.778\times10^{-2}$ &
0.7720 &
$4.779\times10^{-5}$ &
$2.517\times10^{-3}$ \\
\hline
\end{tabular}%
\end{table}

\begin{figure}[H]
    \centering
    \includegraphics[width=0.75\textwidth]{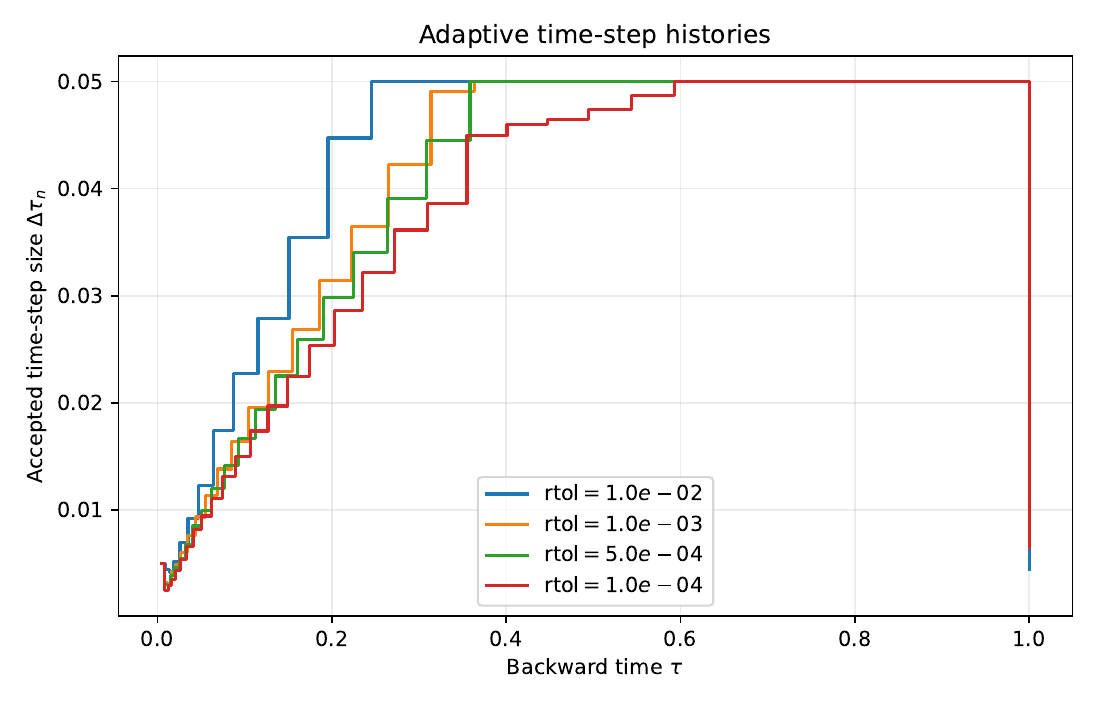}
    \caption{Accepted adaptive time-step sizes for different relative
    tolerances with fixed $\varepsilon=0.04$.}
    \label{fig:adaptive-step-history}
\end{figure}

As $\mathrm{rtol}$ decreases from
$10^{-2}$ to $10^{-4}$, the accepted steps
become smaller and more numerous, while
$E_\infty$ remains near $5\times10^{-5}$.
All accepted non-start-up steps satisfy
the temporal criterion, with at most two
rejected trials per run. This error plateau
is consistent with the finite resolution
of the fixed-penalty reference ($N_t=800$)
and other fixed numerical errors.
Thus, $\eta_n$ measures step acceptance,
not the final global error.
Figure~\ref{fig:adaptive-step-history} shows the accepted step-size
histories. Smaller steps are concentrated near $\tau=0$, where the
payoff is nonsmooth, while progressively larger steps are accepted as
the solution becomes smoother. The controller therefore concentrates
temporal resolution near maturity without maintaining the same small
step throughout the entire pricing interval.
The final decrease in step size is caused by the terminal-step restriction
in \eqref{eq:final-step}, which shortens the last step so that the computation reaches
$\tau=T$ exactly.

\subsection{Dynamic Penalty Behavior}
\label{subsec:dynamic_penalty}

We next examine the dynamic penalty controller independently while keeping the temporal grid fixed. In this experiment, we use $\Delta\tau=0.005, \ N_t=200$ and the baseline penalty-coupling parameters are
$(c_\varepsilon,\alpha)=(1,1)$. For this fixed temporal step, the corresponding time-based baseline
penalty value is set to $\varepsilon^{\mathrm{time}}=0.005$. Since $\gamma_{\mathrm{inc}}=2$, the relaxation mechanism can increase the accepted penalty value to $0.01$ when the measured obstacle violation is sufficiently below its prescribed tolerance.

To investigate the effect of obstacle tolerance, we consider
\[
\mathrm{tol}_{\mathrm{pen}}
\in
\left\{
10^{-2},\ 
5\times10^{-3},\ 
3\times10^{-3}
\right\}.
\]

For each configuration, we record the range and mean value of the
accepted penalty parameter. In Table~\ref{tab:dynamic_penalty},
``Relaxed'' denotes the number and percentage of accepted time steps
for which the accepted penalty value exceeds the baseline value
$0.005$. The quantity $\bar{\varepsilon}$ denotes the mean accepted
penalty value, while ``Switches'' gives the number of changes in the
accepted penalty value between consecutive accepted time steps.
Furthermore,
\[
R_{\mathrm{obs}}^{\max}
=
\max_n R_{\mathrm{obs},n},
\qquad
\zeta_{\max}
=
\max_n \zeta_n,
\]
denote the maximum obstacle residual and the corresponding maximum
normalized obstacle indicator over the computation.

\begin{table}[ht]
\centering
\caption{\small Dynamic penalty behavior for $\Delta\tau=0.005$.
Wall-clock times are reported as mean $\pm$ sample standard deviation
over 10 runs.}
\label{tab:dynamic_penalty}

\footnotesize
\setlength{\tabcolsep}{2.8pt}
\renewcommand{\arraystretch}{1.12}

\begin{tabular}{ccccccccc}
\hline
$\mathrm{tol}_{\mathrm{pen}}$
& $\varepsilon_n$ range
& $\bar{\varepsilon}$
& Relaxed
& $R_{\mathrm{obs}}^{\max}$
& $\zeta_{\max}$
& Switches
& Picard
& Wall time (s) \\
\hline

$10^{-2}$
& $[0.005,\,0.010]$
& $0.00815$
& $126\;(63.0\%)$
& $2.619\times10^{-3}$
& $0.2619$
& 3
& 2429
& $4.36\pm0.40$ \\

$5\times10^{-3}$
& $[0.005,\,0.010]$
& $0.006625$
& $65\;(32.5\%)$
& $2.624\times10^{-3}$
& $0.5248$
& 65
& 2771
& $4.76\pm0.11$ \\

$3\times10^{-3}$
& $[0.005,\,0.010]$
& $0.005325$
& $13\;(6.5\%)$
& $2.627\times10^{-3}$
& $0.8758$
& 26
& 2837
& $4.95\pm0.32$ \\

\hline
\end{tabular}%
\end{table}

As $\mathrm{tol}_{\mathrm{pen}}$ is reduced, relaxation occurs less
frequently. The fraction of accepted time steps using a relaxed
penalty decreases from $63.0\%$ for
$\mathrm{tol}_{\mathrm{pen}}=10^{-2}$ to $6.5\%$ for
$\mathrm{tol}_{\mathrm{pen}}=3\times10^{-3}$. Accordingly,
the mean penalty value moves progressively closer to the stronger
baseline value $0.005$.

For all three tolerance levels, $\zeta_{\max}<1$, showing that the
prescribed obstacle criterion is satisfied throughout the accepted
time steps. No within-step penalty strengthening is required in these
experiments, and no nonlinear Picard failure is observed.

The stricter obstacle requirement nevertheless increases the nonlinear
computational effort. The cumulative number of Picard iterations rises
from 2429 for $\mathrm{tol}_{\mathrm{pen}}=10^{-2}$ to 2837 for
$\mathrm{tol}_{\mathrm{pen}}=3\times10^{-3}$. Consistently, the mean
wall-clock time increases from
$4.36\pm0.40$ seconds to $4.95\pm0.32$ seconds.

These results illustrate the intended behavior of the dynamic penalty
controller. A relatively loose obstacle tolerance permits more frequent
penalty relaxation and therefore weaker penalization, whereas a stricter
tolerance keeps the penalty parameter closer to its baseline value and
requires greater nonlinear computational effort.

\subsection{Comparison of Fixed and Adaptive Strategies}
\label{subsec:strategy-comparison}
We compare four representative strategies: fixed time with fixed penalty
(FT+FP), adaptive time with fixed penalty (AT+FP), fixed time with dynamic
penalty (FT+DP), and adaptive time with dynamic penalty (AT+DP).
The fixed-time computations use $\Delta\tau=0.005$ and $N_t=200$,
whereas the adaptive cases use the baseline temporal parameters of
Table~1. The fixed-penalty cases use $\varepsilon=0.04$, while the
dynamic cases use the baseline penalty-controller parameters.
All errors are measured relative to the working numerical reference
defined in Section~\ref{sec:numerical_experiments}. Each configuration was executed 10 times, and
wall-clock times are reported as mean $\pm$ sample standard deviation.

\begin{table}[H]
\centering
\caption{Accuracy and computational cost of the four numerical strategies.
Wall-clock times are reported as mean $\pm$ sample standard deviation over
10 runs.}
\label{tab:strategy-comparison}

\footnotesize
\setlength{\tabcolsep}{3.5pt}

\begin{tabular}{cccccccc}
\toprule
Method &
$E_\infty$ &
$E_2$ &
$E_{\mathrm{price}}$ &
$R_{\mathrm{obs}}^{\mathrm{final}}$ &
$N_{\mathrm{acc}}/N_{\mathrm{rej}}$ &
Picard &
Wall time (s)\\
\midrule
FT+FP &
$2.219\times10^{-3}$ &
$4.071\times10^{-3}$ &
$1.568\times10^{-1}$ &
$2.266\times10^{-3}$ &
200/0 &
2235 &
$3.90\pm0.18$\\

AT+FP &
$2.226\times10^{-3}$ &
$4.073\times10^{-3}$ &
$1.578\times10^{-1}$ &
$2.272\times10^{-3}$ &
32/2 &
1475 &
$2.85\pm0.27$\\

FT+DP &
$4.627\times10^{-4}$ &
$7.369\times10^{-4}$ &
$2.849\times10^{-2}$ &
$5.091\times10^{-4}$ &
200/0 &
2771 &
$5.31\pm0.64$\\

AT+DP &
$1.674\times10^{-3}$ &
$3.062\times10^{-3}$ &
$1.154\times10^{-1}$ &
$1.721\times10^{-3}$ &
35/7 &
1719 &
$3.23\pm0.06$\\
\bottomrule
\end{tabular}
\end{table}

With a fixed penalty parameter, adaptive time stepping reduces the
accepted-step count from 200 to 32 and the mean wall-clock time from
3.90 s to 2.85 s, while the reported errors remain essentially unchanged.

Among the four strategies, FT+DP gives the smallest reported errors and
final obstacle residual, but also the largest computational cost.
The coupled AT+DP strategy reduces the mean wall-clock time from
5.31 s to 3.23 s relative to FT+DP, corresponding to a reduction of
approximately $39\%$. This saving is accompanied by an increase in the
reported errors: $E_\infty$ is approximately 3.6 times larger and
$E_{\mathrm{price}}$ approximately 4.1 times larger than for FT+DP.
Accordingly, this comparison should be interpreted as a cost-accuracy
trade-off rather than as an efficiency improvement at matched accuracy.

Nevertheless, AT+DP remains more accurate and gives a smaller final
obstacle residual than FT+FP, while requiring substantially fewer Picard
iterations than FT+DP. These results illustrate the intended trade-off
between accuracy, obstacle enforcement, and computational cost.

As an additional spatial-resolution check, the FT+FP and AT+FP
computations were repeated on the refined grid
$(N_x,N_v)=(200,100)$, using a corresponding working reference
computed on the same grid with $N_t=3200$ and
$\varepsilon=3.125\times10^{-4}$. The corresponding maximum-norm
errors were $2.239\times10^{-3}$ and $2.246\times10^{-3}$,
respectively, while the discrete $L^2$ and selected-price errors also
remained nearly identical. The adaptive computation required 35
accepted steps, compared with 200 for the fixed-time calculation.
Thus, the main accuracy comparison between FT+FP and AT+FP is
preserved under this spatial refinement.

\subsection{Sensitivity to the Penalty-Coupling Parameters}
\label{subsec:penalty-sensitivity}

We examine the sensitivity of the coupled method to the parameters
$c_\varepsilon$ and $\alpha$. The baseline configuration is
$(c_\varepsilon,\alpha)=(1,1)$. We vary
$\alpha\in\{0.75,1,1.05\}$ with $c_\varepsilon=1$, and
$c_\varepsilon\in\{0.75,1,1.5\}$ with $\alpha=1$.
All remaining parameters are kept at their baseline values, and all
reported errors are measured relative to the working numerical reference
defined in Section~\ref{subsec:experimental-setup}. The results are summarized in
Table~\ref{tab:penalty-sensitivity}.

\begin{table}[H]
\centering
\caption{Sensitivity of the coupled adaptive method to
$c_\varepsilon$ and $\alpha$.}
\label{tab:penalty-sensitivity}
\footnotesize
\setlength{\tabcolsep}{3.2pt}

\begin{tabular}{cccccccc}
\toprule
$c_\varepsilon$ & $\alpha$ &
$E_\infty$ & $E_2$ & $E_{\mathrm{price}}$ &
$R_{\mathrm{obs}}^{\mathrm{final}}$ &
$N_{\mathrm{acc}}/N_{\mathrm{rej}}$ & Picard\\
\midrule
1.00 & 0.75 &
$4.892\times10^{-3}$ &
$9.588\times10^{-3}$ &
$3.681\times10^{-1}$ &
$4.937\times10^{-3}$ &
32/2 & 1348\\

1.00 & 1.00 &
$1.674\times10^{-3}$ &
$3.062\times10^{-3}$ &
$1.154\times10^{-1}$ &
$1.721\times10^{-3}$ &
35/7 & 1719\\

1.00 & 1.05 &
$1.925\times10^{-3}$ &
$3.488\times10^{-3}$ &
$1.339\times10^{-1}$ &
$1.971\times10^{-3}$ &
35/7 & 2069\\

0.75 & 1.00 &
$2.073\times10^{-3}$ &
$3.753\times10^{-3}$ &
$1.457\times10^{-1}$ &
$2.119\times10^{-3}$ &
34/7 & 2343\\

1.50 & 1.00 &
$3.188\times10^{-3}$ &
$6.157\times10^{-3}$ &
$2.355\times10^{-1}$ &
$3.234\times10^{-3}$ &
34/6 & 1472\\
\bottomrule
\end{tabular}
\end{table}

Among the tested configurations, the baseline choice
$(c_\varepsilon,\alpha)=(1,1)$ gives the smallest reported errors and
final obstacle residual. For $\alpha=0.75$, the obstacle criterion
triggers two within-step penalty-correction updates, showing that the
strengthening mechanism of Section~\ref{subsec:within-step-penalty} is activated when required.
Increasing $\alpha$ to $1.05$ raises the cumulative Picard count from
1719 to 2069 without improving the reported accuracy. Changing
$c_\varepsilon$ away from its baseline value likewise alters the balance
between accuracy, obstacle enforcement, and nonlinear computational
effort. Therefore, $c_\varepsilon$ and $\alpha$ should be regarded as
numerical tuning parameters rather than universal constants.

 \subsection{Additional Test under Increased Jump Intensity}

To provide an additional check of the coupled controller beyond the
baseline Bates parameter regime, we consider a higher jump intensity
$\lambda=0.20$, while keeping the remaining model and numerical
parameters unchanged. For each regime, the errors are measured relative
to a separate working numerical reference computed with
$N_t=3200$ and $\varepsilon=3.125\times10^{-4}$ on the fixed spatial
grid $(N_x,N_v)=(100,50)$.

\begin{table}[H]
\centering
\caption{Performance of the coupled AT+DP method for the baseline and
high-jump Bates regimes.}
\label{tab:bates-robustness}

\footnotesize
\setlength{\tabcolsep}{3.5pt}

\begin{tabular}{cccccccc}
\toprule
Regime &
$\lambda$ &
$E_\infty$ &
$E_2$ &
$E_{\mathrm{price}}$ &
$R_{\mathrm{obs}}^{\mathrm{final}}$ &
$N_{\mathrm{acc}}/N_{\mathrm{rej}}$ &
Picard\\
\midrule

Baseline &
0.05 &
$1.674\times10^{-3}$ &
$3.062\times10^{-3}$ &
$1.154\times10^{-1}$ &
$1.721\times10^{-3}$ &
35/7 &
1719\\

High jump &
0.20 &
$2.691\times10^{-3}$ &
$4.780\times10^{-3}$ &
$1.730\times10^{-1}$ &
$2.562\times10^{-3}$ &
34/7 &
1689\\

\bottomrule
\end{tabular}
\end{table}

Increasing the jump intensity from $\lambda=0.05$ to $\lambda=0.20$
leads to larger reported errors and obstacle residual, as expected for
the more demanding regime. Nevertheless, the adaptive controller remains
operational with a comparable number of accepted and rejected trials
($34/7$ versus $35/7$), no nonlinear solver failures, and a similar
cumulative Picard count. Moreover, the maximum accepted temporal and
obstacle indicators remain below one. This experiment provides an
additional robustness check of the coupled procedure under a substantially
higher jump intensity, although it does not constitute a comprehensive
parameter-space study.

\subsection{Sensitivity to Spatial Boundary Treatments}
\label{sec:variance_boundary_sensitivity}

We examined the sensitivity of the reported American
put prices to numerical treatment of the variance boundaries.
At $v=0$, we compared the homogeneous Neumann closure used
in the original computations with a diagnostic closure
retaining the limiting Bates PIDE at the degenerate boundary.
For each $N_v$, both methods used the same spatial grid,
with $N_x=100$, $v_{\max}=1$, and a common history of
35 accepted time steps and penalty parameters recorded
from the baseline adaptive computation at $N_v=50$.
These replayed computations are separate from the
results reported in Tables~~\ref{tab:reference-refinement}--~\ref{tab:variance_boundary_sensitivity}.

\begin{table}[htbp]
\centering
\caption{Sensitivity of selected American
put prices ($K=100$, $v=0.04$) to variance-boundary treatment.
Panel (a): $N_x=100$, $v_{\max}=1$, with the same 35-step
temporal and penalty history in both lower-boundary
treatments. Panel (b): the original homogeneous Neumann
closure, with $\Delta v=0.02$ and the same temporal and
penalty history for all three upper variance limits.}
\label{tab:variance_boundary_sensitivity}

\small
\begin{tabular}{rrrr}
\hline
\multicolumn{4}{l}{
\textbf{(a) Lower variance boundary:} $V(100,0.04)$}\\
$N_v$ & Neumann & Limiting PIDE & Absolute difference\\
\hline
25  & 6.390226 & 5.878086 & 0.512140\\
50  & 6.019171 & 5.908047 & 0.111124\\
75  & 5.957805 & 5.906764 & 0.051041\\
100 & 5.934793 & 5.905601 & 0.029192\\
150 & 5.919447 & 5.905957 & 0.013490\\
\hline

\multicolumn{4}{l}{
\textbf{(b) Upper variance boundary:} Neumann closure}\\
$v_{\max}$ ($N_v$) & $V(80,0.04)$
& $V(100,0.04)$ & $V(120,0.04)$\\
\hline
0.5 (25) & 19.883410906 & 6.019170689 & 1.883476534\\
1.0 (50) & 19.883410909 & 6.019170785 & 1.883476605\\
1.5 (75) & 19.883410909 & 6.019170853 & 1.883476656\\
\hline
\end{tabular}
\end{table}

In panel (a), the difference between the two
lower-boundary closures decreases with refinement in the
variance direction, but it is appreciable on the baseline
grid. All replayed non-start-up trials satisfied the
temporal criterion $\eta_n\leq1$, and the obstacle
criterion $\zeta_n\leq1$ was also satisfied.
Panel (b) isolates the upper-boundary location by keeping
$\Delta v$ and the time/penalty history fixed: across the
three tested domains, the largest difference among the
three selected option prices is below $1.7\times10^{-7}$.
These findings quantify finite-grid boundary sensitivity
in the tested parameter regime; they neither establish
the exact Bates option value nor prove convergence or
unconditional stability of the full scheme.

Left log-price boundary.
We varied $x_{\min}\in\{-4,-5,-6\}$ at fixed
$\Delta x=0.1$ ($N_x=90,100,110$),
$x_{\max}=5$, and $N_v=50$, comparing
$u=1$ with $u=1-e^{x_{\min}}$ using the
same payoff-based jump extension.
All six AT+DP runs had 35 accepted and
7 rejected steps. At $v=0.04$, prices
at $S=80,100,120$ agreed within
$1.1\times10^{-13}$; in particular,
$V(100,0.04)=6.01917078$.
The boundary mismatch $e^{x_{\min}}$
remains, so these results do not establish
a uniform truncation-error bound.

\subsection{Stability Diagnostics for the Explicit Jump Treatment}
\label{subsec:jump-stability-diagnostic}

We assessed the stability restrictions of the explicit
variable-step AB2 jump treatment in \eqref{eq:variable-ab2} by
combining a linearized spectral diagnostic with numerical runs of
the full adaptive solver. The jump matrix was assembled from the
same Simpson quadrature and compact first-derivative operator as
the implementation, using $N_x=100$ and homogeneous boundary
perturbations. For each eigenvalue $\mu$ of the discrete jump
matrix, the roots of the scalar AB2 recurrence were checked at
fixed values of the proposed step $h$ and ratio $\chi$.
For the baseline $\lambda=0.05$, $h=0.05$ and $\chi=2$, all
frozen-ratio modal roots lie within the unit disk. By contrast,
with $\lambda=4$, $h=0.20$, and $\chi=2$, the largest root
modulus is approximately $1.385$. These are diagnostics of the
isolated jump treatment, not stability bounds for the full
splitting or for arbitrary variable-step histories.

\begin{table}[htbp]
\centering
\caption{Adaptive-solver sensitivity to jump intensity and
maximum step size on the diagnostic spatial grid $(N_x,N_v)=(48,24)$.
Other numerical parameters follow Table~\ref{tab:baseline_parameters}.
Here $h_{\mathrm{used,max}}$ and $\chi_{\mathrm{used,max}}$ denote maxima
among the accepted steps, excluding the start-up ratio.}
\label{tab:jump-stability-diagnostic}

\footnotesize
\setlength{\tabcolsep}{5pt}
\begin{tabular}{cccccc}
\hline
$\lambda$ & $h_{\max}$ & $N_{\mathrm{acc}}$ &
$h_{\mathrm{used,max}}$ & $\chi_{\mathrm{used,max}}$ & $\zeta_{\max}$ \\
\hline
$0.05$ & $0.025$ & 51 & $0.0250$ & $1.415$ & $0.333$ \\
$0.05$ & $0.050$ & 35 & $0.0500$ & $1.415$ & $0.581$ \\
$0.05$ & $0.200$ & 29 & $0.1334$ & $1.824$ & $0.973$ \\
$0.20$ & $0.025$ & 51 & $0.0250$ & $1.392$ & $0.332$ \\
$0.20$ & $0.050$ & 35 & $0.0500$ & $1.392$ & $0.576$ \\
$0.20$ & $0.200$ & 28 & $0.1301$ & $1.745$ & $0.978$ \\
$1.00$ & $0.025$ & 51 & $0.0250$ & $1.389$ & $0.328$ \\
$1.00$ & $0.050$ & 34 & $0.0500$ & $1.389$ & $0.554$ \\
$1.00$ & $0.200$ & 28 & $0.1297$ & $1.389$ & $0.994$ \\
\hline
\end{tabular}
\end{table}
Table~\ref{tab:jump-stability-diagnostic} reports nine representative
configurations from the 16 adaptive runs.
All tested adaptive runs completed without nonlinear-solver
failure, and their accepted steps satisfied the prescribed temporal
and obstacle criteria. With $h_{\max}=0.20$, however, the
controller selected steps no larger than approximately $0.13$.
As a separate ratio diagnostic, fixed-penalty calculations with
40 prescribed steps and alternating step-size ratios of $2$ and
$1/2$ (or $3$ and $1/3$) also completed for
$\lambda\in\{0.05,0.20,1.00\}$. These observations indicate
practical robustness in the tested settings but do not prove
unconditional stability, nor do they establish that the empirical
cap is sufficient for every jump intensity or spatial resolution.
\subsection{Robustness with Respect to Model Parameters}
\label{subsec:model-parameter-robustness}

To complement the penalty-parameter sensitivity in Section~\ref{subsec:penalty-sensitivity}
and the jump-intensity comparison in Table~\ref{tab:bates-robustness}, we varied the remaining model
parameters of Table~\ref{tab:baseline_parameters} one at a time.
The tests use the original coupled AT+DP algorithm on the baseline
spatial grid $(N_x,N_v)=(100,50)$, keeping all numerical controls
unchanged. Each model parameter was examined at two additional
values, and maturity was also varied. In all model-parameter cases
the Feller condition $2\kappa\theta>\sigma^2$ holds.

Because the transformed model uses $x=\log(S/K)$ and $u=V/K$,
changing the strike $K$ at fixed moneyness only rescales the monetary price;
it is therefore not varied independently in the normalized model.

\begin{table}[H]
\centering
\caption{One-at-a-time parameter-robustness checks for the coupled AT+DP solver.
All cases use $(N_x,N_v)=(100,50)$. $R_{\mathrm{obs}}^{\mathrm{final}}$ is
the final obstacle shortfall, and $\zeta_{\max}$ is the largest accepted
obstacle indicator. Parameter values not shown remain at Table~\ref{tab:baseline_parameters}.}
\label{tab:model-parameter-robustness}

\footnotesize
\setlength{\tabcolsep}{4pt}
\renewcommand{\arraystretch}{1.05}
\begin{tabular}{lrrrrr}
\toprule
Parameter & Value & $V(100)$ & $R_{\mathrm{obs}}^{\mathrm{final}}$ & $N_{\mathrm{acc}}/N_{\mathrm{rej}}$ & $\zeta_{\max}$ \\
\midrule
Baseline & -- & 6.01917 & $1.721\times10^{-3}$ & 35/7 & 0.565 \\
$r$ & 0.030 & 6.60696 & $9.062\times10^{-4}$ & 35/7 & 0.363 \\
$r$ & 0.080 & 5.26166 & $2.750\times10^{-3}$ & 35/8 & 0.842 \\
$\kappa$ & 1.500 & 6.01916 & $2.026\times10^{-3}$ & 35/7 & 0.612 \\
$\kappa$ & 3.000 & 6.02014 & $1.923\times10^{-3}$ & 35/7 & 0.548 \\
$\theta$ & 0.030 & 5.58270 & $1.714\times10^{-3}$ & 35/7 & 0.579 \\
$\theta$ & 0.060 & 6.89129 & $1.508\times10^{-3}$ & 35/7 & 0.553 \\
$\sigma$ & 0.200 & 5.97509 & $2.254\times10^{-3}$ & 34/7 & 0.649 \\
$\sigma$ & 0.380 & 6.11349 & $1.856\times10^{-3}$ & 35/7 & 0.556 \\
$\rho$ & -0.800 & 6.03259 & $1.767\times10^{-3}$ & 36/7 & 0.571 \\
$\rho$ & 0.000 & 5.93118 & $1.619\times10^{-3}$ & 33/7 & 0.583 \\
$m_J$ & -0.100 & 6.05331 & $2.594\times10^{-3}$ & 34/7 & 0.566 \\
$m_J$ & 0.000 & 6.00187 & $1.811\times10^{-3}$ & 35/7 & 0.563 \\
$\delta_J$ & 0.080 & 6.00147 & $1.765\times10^{-3}$ & 35/7 & 0.566 \\
$\delta_J$ & 0.160 & 6.07709 & $2.564\times10^{-3}$ & 34/7 & 0.561 \\
\midrule
$T$ & 0.50 & 4.38809 & $1.702\times10^{-3}$ & 25/7 & 0.511 \\
$T$ & 1.50 & 7.09882 & $1.474\times10^{-3}$ & 45/7 & 0.565 \\
\bottomrule
\end{tabular}
\end{table}

Table~\ref{tab:model-parameter-robustness}
summarizes the results of the parameter-robustness
tests. 
For the parameter variations with $T=1$, the
accepted-step count ranged from 33 to 36;
the two maturity tests required 25 and
45 accepted steps, respectively.
The maximum accepted temporal indicator was
below $0.999$. The changes in $V(100)$ represent sensitivity of option prices
to the model parameters and are not estimates of numerical error.
The results provide empirical evidence of controller robustness over
the tested ranges, but do not establish uniform pricing accuracy,
convergence, or stability for arbitrary parameter combinations.

To assess variance-grid sensitivity
in the $r=0.08$ regime, we performed a
fixed-step, fixed-penalty comparison with
$(N_x,N_t)=(100,200)$, $\varepsilon=0.005$,
and $v_{\max}=1$. Increasing $N_v$ from
$25$ to $150$ reduced the difference in
$V(100,0.04)$ between the Neumann and
diagnostic limiting-PIDE closures from
$0.46243$ to $0.01079$. The maximum
obstacle shortfall at $v=0$ under the
Neumann closure was nonmonotone, reaching
$3.16\times10^{-3}$ at $N_v=50$ but
remaining below $4.3\times10^{-4}$ for
$N_v=75,100,150$. These observations
indicate sensitivity to variance-grid
resolution but do not establish convergence
to the exact American option solution.

\subsection{Discrete PIDE Complementarity Diagnostics}
\label{subsec:complementarity-diagnostics}

The obstacle shortfall $R_{\mathrm{obs}}$ alone does not assess the
PIDE supersolution inequality or the complementarity condition. For an
additional final-time diagnostic, let $\mathcal I_h$ denote the interior
spatial nodes, where the discrete PIDE is evaluated rather than replaced
by a boundary equation. Using the two half-step states from the last
accepted adaptive interval $[\tau_n,\tau_{n+1}]$, we define
\begin{equation}
\label{eq:complementarity-diagnostics}
\begin{aligned}
F_h^{n+1} & = T\left[
 \frac{3U_H^{n+1}-4U_H^{n+\frac12}+U^n}{h_n}
 -\mathcal L_h U_H^{n+1}-\mathcal J_h U_H^{n+1}\right],\\
R_{\mathrm{PIDE}}^{-} & = \|(-F_h^{n+1})_+\|_{\infty,\mathcal I_h},\\
R_{\mathrm{VI}} & = \|\min(U_H^{n+1}-\widetilde g,F_h^{n+1})\|_{\infty,\mathcal I_h},\\
R_{\times} & = \|(U_H^{n+1}-\widetilde g)_+(F_h^{n+1})_+\|_{\infty,\mathcal I_h}.
\end{aligned}
\end{equation}
The minimum and products in \eqref{eq:complementarity-diagnostics}
are evaluated componentwise. Multiplication by $T$ gives the PIDE residual
the same normalized value scale as the obstacle gap (all tests below have
$T=1$). In particular, $R_{\mathrm{VI}}=0$ is equivalent to the discrete
obstacle inequality, nonnegative PIDE residual, and their complementary
slackness at the interior nodes. The time derivative above is reconstructed
only for postprocessing; it is not the temporal formula used to advance
the Douglas--AB2 solver. The jump and local spatial operators are evaluated
at the final state, while $R_{\mathrm{obs}}$ remains evaluated on the full
spatial grid, including the variance boundaries.

\begin{table}[H]
\centering
\caption{Final-time complementarity diagnostics for AT+DP on
$(N_x,N_v)=(100,50)$ with $T=1$. The PIDE and complementarity quantities
are measured over interior spatial nodes; $R_{\mathrm{obs}}$ is measured
over the entire grid.}
\label{tab:complementarity-diagnostics}

\footnotesize
\setlength{\tabcolsep}{3.5pt}
\renewcommand{\arraystretch}{1.08}
\begin{tabular}{ccrrrr}
\toprule
$\lambda$ & $\mathrm{rtol}$ & $R_{\mathrm{obs}}$ & $R_{\mathrm{PIDE}}^{-}$ & $R_{\mathrm{VI}}$ & $R_{\times}$ \\
\midrule
$0.05$ & $10^{-2}$ & $2.449\times10^{-3}$ & $2.495\times10^{-4}$ & $2.204\times10^{-3}$ & $2.180\times10^{-6}$ \\
$0.05$ & $10^{-3}$ & $1.721\times10^{-3}$ & $1.353\times10^{-3}$ & $1.816\times10^{-3}$ & $4.950\times10^{-6}$ \\
$0.05$ & $10^{-4}$ & $2.096\times10^{-3}$ & $6.320\times10^{-4}$ & $1.859\times10^{-3}$ & $3.574\times10^{-6}$ \\
$0.20$ & $10^{-3}$ & $2.562\times10^{-3}$ & $1.999\times10^{-4}$ & $2.350\times10^{-3}$ & $2.364\times10^{-6}$ \\
\bottomrule
\end{tabular}
\end{table}

The additional diagnostics in Table~\ref{tab:complementarity-diagnostics} are nonzero even when accepted steps satisfy the obstacle tolerance. Their values depend on the chosen temporal
reconstruction and on the fixed spatial discretization; in particular,
tightening $\mathrm{rtol}$ does not monotonically reduce the final-time
complementarity residual because the adaptive penalty and the final-step
history also change. The tests therefore assess the extent of discrete
complementarity violation in the reported regimes, without asserting exact
complementarity or convergence of the complete scheme.
\subsection{Accuracy--Cost Comparison at Common Error Thresholds}
\label{subsec:matched-accuracy}

The baseline comparison in Table~\ref{tab:strategy-comparison}
shows a reduction in computation time for AT+DP relative to FT+DP,
but with larger numerical errors. To assess computational cost at
comparable accuracy, we compare the two methods on the same
$(N_x,N_v)=(100,50)$ spatial grid, using the baseline Bates parameters
and the common fixed-grid working reference of
Section~\ref{subsec:experimental-setup}. The dynamic-penalty rule
is unchanged. For FT+DP we vary the number of uniform time steps
$N_t$; for AT+DP we vary the upper step-size bound $h_{\max}$ while
keeping $\mathrm{rtol}=10^{-3}$. The measured errors are differences
from the working numerical reference, not errors relative to an
exact American option solution.

We select three illustrative shared error budgets, specified
as upper bounds on $(E_\infty,E_{\mathrm{price}})$:
$(1.3\times10^{-3},0.080)$, $(5.5\times10^{-4},0.030)$, and
$(3.5\times10^{-4},0.020)$. Both methods satisfy both bounds in
each corresponding pair, although their errors are not identical.
Wall-clock time is measured for the solver, including nonlinear
iterations and rejected adaptive trials but excluding solver
initialization and reference-error postprocessing. Each configuration
was timed 10 times on the ASUS ExpertBook described in
Section~\ref{subsec:experimental-setup}; we report the sample mean
and standard deviation.

\begin{table}[H]
\centering
\caption{Accuracy and wall-clock cost of FT+DP and AT+DP at
three common error budgets on $(N_x,N_v)=(100,50)$.
Times are the sample mean $\pm$ standard deviation of 10 runs
on the same laptop.}
\label{tab:matched-accuracy}

\footnotesize
\setlength{\tabcolsep}{3.5pt}
\renewcommand{\arraystretch}{1.08}
\begin{tabular}{lllrrrrr}
\toprule
Budget & Method & Setting & $E_\infty$ & $E_{\mathrm{price}}$ & $N_{\mathrm{acc}}$ & Picard & Time (s) \\
\midrule
Coarse & FT+DP & $N_t=50$ & $1.221\times10^{-3}$ & 0.0792 & 50 & 810 & $2.13\pm0.12$ \\
Coarse & AT+DP & $h_{\max}=0.025$ & $1.051\times10^{-3}$ & 0.0648 & 51 & 2298 & $7.13\pm0.76$ \\
\midrule
Medium & FT+DP & $N_t=150$ & $4.874\times10^{-4}$ & 0.0297 & 150 & 2163 & $5.01\pm0.17$ \\
Medium & AT+DP & $h_{\max}=0.010$ & $5.003\times10^{-4}$ & 0.0276 & 107 & 4221 & $10.44\pm1.40$ \\
\midrule
Fine & FT+DP & $N_t=400$ & $3.142\times10^{-4}$ & 0.0185 & 400 & 4535 & $10.13\pm0.21$ \\
Fine & AT+DP & $h_{\max}=0.0075$ & $3.372\times10^{-4}$ & 0.0173 & 139 & 5317 & $18.54\pm8.21$ \\
\bottomrule
\end{tabular}
\end{table}

\begin{figure}[H]
\centering
\includegraphics[width=0.82\linewidth]{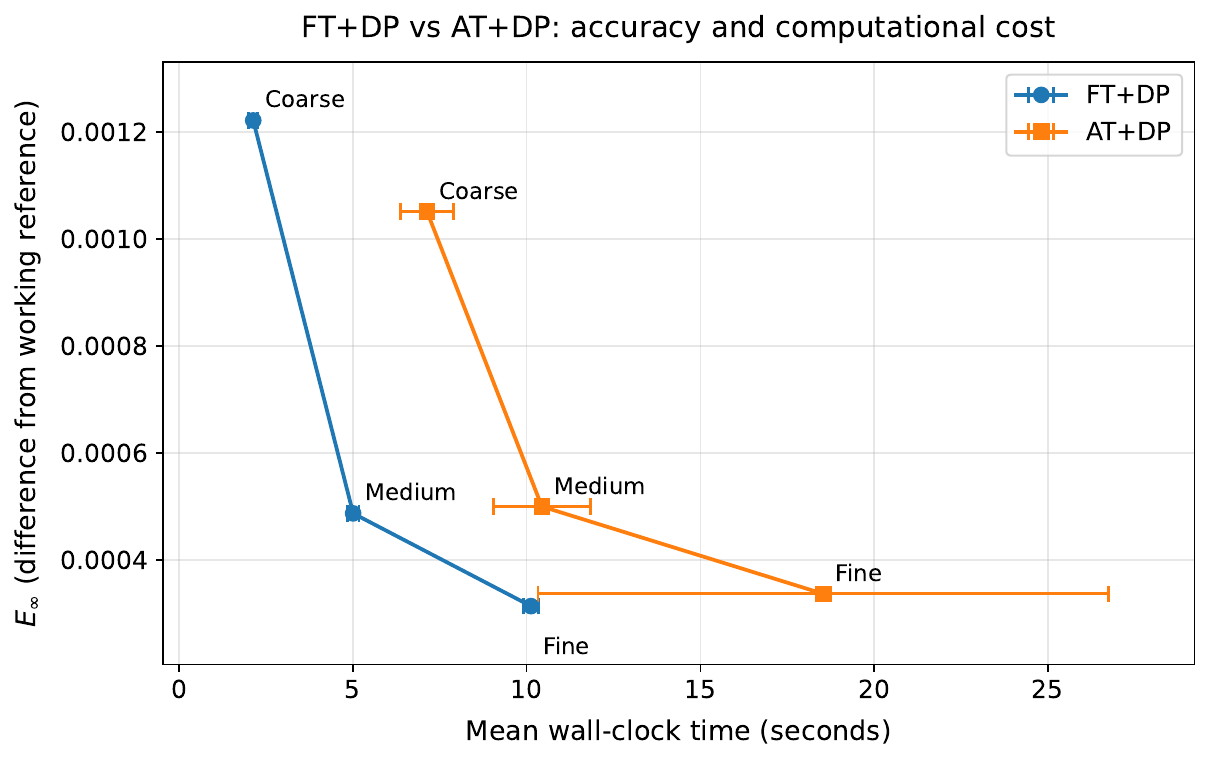}
\caption{Accuracy--cost comparison for FT+DP and AT+DP
at the shared error budgets of Table~\ref{tab:matched-accuracy}.
The vertical axis shows the normalized maximum-norm difference
$E_\infty$ from the common working reference. Horizontal error
bars show one sample standard deviation of wall-clock time across
10 runs; they do not represent uncertainty in $E_\infty$.}
\label{fig:matched-accuracy}
\end{figure}

Each adaptive trial requires one full
step and two half steps, with two directional
ADI stages per Picard iteration. The larger
cumulative Picard counts of AT+DP in
Table~\ref{tab:matched-accuracy} help explain
its higher computational cost, although the
individual overhead components have not
been separately timed.

Table~\ref{tab:matched-accuracy} and
Figure~\ref{fig:matched-accuracy} show that FT+DP has a lower
mean wall-clock time at all three tested error budgets. In
particular, at the medium budget the times are
$5.01\pm0.17$~s for FT+DP and $10.44\pm1.40$~s for AT+DP,
with comparable $E_\infty$. The fine-budget AT+DP measurements
have substantial variability ($18.54\pm8.21$~s), which is retained
rather than discarding slower runs. Thus, the baseline time saving
in Table~\ref{tab:strategy-comparison} should be interpreted as
a cost--accuracy trade-off, not an improvement at matched accuracy.
This conclusion concerns the tested spatial grid, model parameters,
error budgets, and numerical reference; it does not establish a
universal efficiency ranking.

\section{Conclusion}
\label{sec:conclusion}

This paper developed a coupled adaptive framework for pricing
American put options under the Bates stochastic-volatility
jump-diffusion model. The method combines adaptive time stepping
with dynamic penalty control while keeping the spatial
discretization fixed. With a fixed penalty, the number of accepted
steps decreases from 200 to 32 with nearly unchanged reported
errors. The fully coupled AT+DP method reduces baseline
wall-clock time by approximately 39\% relative to FT+DP, but
increases numerical error. At three common error budgets,
FT+DP has lower mean computational cost, indicating a
cost--accuracy trade-off rather than an efficiency improvement
at comparable accuracy. The sensitivity, higher-jump, and
spatial-refinement tests further indicate that the coupled
controller remains operational beyond the baseline configuration.
These experiments do not constitute a comprehensive
parameter-space study. Further work may consider broader
Bates parameter regimes and adaptive spatial refinement.


\section{Acknowledgments}
The authors gratefully acknowledge the contribution of the Slovak Research and Development Agency VEGA 1-0493-24 (D.\v{S}).

\bibliographystyle{siam}

\bibliography{paper}

@article{Bates1996,
  author  = {Bates, David S.},
  title   = {Jumps and Stochastic Volatility: Exchange Rate Processes Implicit in Deutsche Mark Options},
  journal = {The Review of Financial Studies},
  volume  = {9},
  number  = {1},
  pages   = {69--107},
  year    = {1996}
}

@article{Andersen2000,
  author  = {Andersen, Leif and Andreasen, Jesper},
  title   = {Jump-Diffusion Processes: Volatility Smile Fitting and Numerical Methods for Option Pricing},
  journal = {Review of Derivatives Research},
  volume  = {4},
  pages   = {231--262},
  year    = {2000}
}

@article{DHalluin2004,
  author  = {d'Halluin, Yann and Forsyth, Peter A. and Labahn, George},
  title   = {A Penalty Method for American Options with Jump Diffusion Processes},
  journal = {Numerische Mathematik},
  volume  = {97},
  pages   = {321--352},
  year    = {2004}
}

@article{IntHout2010,
  author  = {in 't Hout, Karel J. and Foulon, Steven},
  title   = {ADI Finite Difference Schemes for Option Pricing in the Heston Model with Correlation},
  journal = {International Journal of Numerical Analysis and Modeling},
  volume  = {7},
  pages   = {303--320},
  year    = {2010}
}

@article{Salmi2014,
  author  = {Salmi, Sami and Toivanen, Jari},
  title   = {IMEX Schemes for Pricing Options under Jump-Diffusion Models},
  journal = {Applied Numerical Mathematics},
  volume  = {84},
  pages   = {33--45},
  year    = {2014}
}

@article{Zvan1998,
  author  = {Zvan, Renata and Forsyth, Peter A. and Vetzal, Kenneth R.},
  title   = {Penalty Methods for American Options with Stochastic Volatility},
  journal = {Journal of Computational and Applied Mathematics},
  volume  = {91},
  pages   = {199--218},
  year    = {1998}
}

@article{Clevenhaus2020,
  author  = {Clevenhaus, Andreas and Ehrhardt, Matthias and G{\"u}nther, Michael and {\v S}ev{\v c}ovi{\v c}, Daniel},
  title   = {Pricing American Options with a Non-Constant Penalty Parameter},
  journal = {Journal of Risk and Financial Management},
  volume  = {13},
  number  = {6},
  pages   = {124},
  year    = {2020},
  }

@article{Soderlind2002,
  author  = {S{\"o}derlind, Gustaf},
  title   = {Automatic Control and Adaptive Time-Stepping},
  journal = {Numerical Algorithms},
  volume  = {31},
  number  = {1--4},
  pages   = {281--310},
  year    = {2002},
}

@article{Persson2010,
  author  = {Persson, Jonas and von Sydow, Lina},
  title   = {Pricing American Options Using a Space-Time Adaptive Finite Difference Method},
  journal = {Mathematics and Computers in Simulation},
  volume  = {80},
  number  = {9},
  pages   = {1922--1935},
  year    = {2010},
  }

@article{VonSydow2015,
  author  = {von Sydow, Lina and Toivanen, Jari and Zhang, Chi},
  title   = {Adaptive Finite Differences and IMEX Time-Stepping to Price Options under Bates Model},
  journal = {International Journal of Computer Mathematics},
  volume  = {92},
  number  = {12},
  pages   = {2515--2529},
  year    = {2015},
}

@article{ForsythVetzal2002,
  author  = {Forsyth, Peter A. and Vetzal, Kenneth R.},
  title   = {Quadratic Convergence for Valuing American Options Using a Penalty Method},
  journal = {SIAM Journal on Scientific Computing},
  volume  = {23},
  number  = {6},
  pages   = {2095--2122},
  year    = {2002},
}

@article{Wang2006,
  author  = {Wang, S. and Yang, Xiaoqi and Teo, Kok Lay},
  title   = {Power Penalty Method for a Linear Complementarity Problem Arising from American Option Valuation},
  journal = {Journal of Optimization Theory and Applications},
  volume  = {129},
  number  = {2},
  pages   = {227--254},
  year    = {2006},
  }

@article{Zhang2008,
  author  = {Zhang, Kai and Yang, Xiaoqi and Teo, Kok Lay},
  title   = {Convergence Analysis of a Monotonic Penalty Method for American Option Pricing},
  journal = {Journal of Mathematical Analysis and Applications},
  volume  = {348},
  number  = {2},
  pages   = {915--926},
  year    = {2008},
  }

@article{Arevalo2021,
  author  = {Ar{\'e}valo, Carmen and S{\"o}derlind, Gustaf
             and Hadjimichael, Yiannis and Fekete, Imre},
  title   = {Local Error Estimation and Step Size Control in Adaptive
             Linear Multistep Methods},
  journal = {Numerical Algorithms},
  volume  = {86},
  pages   = {537--563},
  year    = {2021},
  }

@incollection{BagheriRenaniSevcovic2026,
  author    = {Bagheri Renani, Neda and {\v{S}}ev{\v{c}}ovi{\v{c}}, Daniel},
  title     = {Finite Element Solution of the Two-Dimensional Bates Model for Option Pricing Under Stochastic Volatility and Jumps},
  booktitle = {New Trends in the Applications of Differential Equations in Sciences},
  editor    = {Slavova, Angela},
  series    = {Springer Proceedings in Mathematics \& Statistics},
  volume    = {533},
  pages     = {325--336},
  publisher = {Springer Nature Switzerland},
  year      = {2026},
}

@article{BagheriHaghighi2017,
  author  = {Bagheri, Neda and Haghighi, Hassan Karnameh},
  title   = {A Comparison Study of ADI and LOD Methods on Option Pricing Models},
  journal = {Journal of Mathematical Finance},
  volume  = {7},
  number  = {2},
  pages   = {275--290},
  year    = {2017},
}

@article{Stamicar1999,
  author  = {Stamicar, Robert and {\v{S}}ev{\v{c}}ovi{\v{c}}, Daniel and Chadam, John},
  title   = {The Early Exercise Boundary for the American Put Near Expiry: Numerical Approximation},
  journal = {Canadian Applied Mathematics Quarterly},
  volume  = {7},
  number  = {4},
  pages   = {427--444},
  year    = {1999}
}

@article{LaukoSevcovic2010,
  author  = {Lauko, Martin and {\v{S}}ev{\v{c}}ovi{\v{c}}, Daniel},
  title   = {Comparison of Numerical and Analytical Approximations of the Early Exercise Boundary of American Put Options},
  journal = {The ANZIAM Journal},
  volume  = {51},
  number  = {4},
  pages   = {430--448},
  year    = {2010},
}

@article{Merton1976JFE,
  author  = {Merton, Robert C.},
  title   = {Option pricing when underlying stock returns are discontinuous},
  journal = {Journal of Financial Economics},
  volume  = {3},
  pages   = {125--144},
  year    = {1976}
}

@article{Heston1993,
  author  = {Heston, Steven L.},
  title   = {A Closed-Form Solution for Options with Stochastic Volatility with Applications to Bond and Currency Options},
  journal = {The Review of Financial Studies},
  volume  = {6},
  number  = {2},
  pages   = {327--343},
  year    = {1993}
}

@book{ContTankov2004Book,
  author    = {Cont, Rama and Tankov, Peter},
  title     = {Financial Modelling with Jump Processes},
  publisher = {Chapman \& Hall/CRC},
  address   = {Boca Raton, FL},
  year      = {2004}
}

\end{document}